\documentclass[letterpaper,longbibliography,reprint,aps,prapplied]{revtex4-2}

\usepackage{graphicx}
\usepackage{dcolumn}
\usepackage{bm}
\usepackage{amsmath}
\usepackage{xcolor}
\usepackage{soul}
\setstcolor{red}

\begin{document}
\preprint{APS/123-QED}

\title{Modeling conductive thermal transport in three-dimensional \\fibrous media with fiber-to-fiber contacts}

\author{Clémence Gaunand}
\affiliation{
 Institut Langevin, ESPCI Paris, Université PSL, CNRS - Paris (France),
}%
\affiliation{
 Saint-Gobain Research Paris - Aubervilliers (France)
}%
\affiliation{
 Institut Pprime, CNRS, Université de Poitiers, ISAE-ENSMA - Poitiers (France)
}%

\author{Yannick De Wilde}
\affiliation{
 Institut Langevin, ESPCI Paris, Université PSL, CNRS - Paris (France),
}%

\author{Adrien François}
\affiliation{
 Saint-Gobain Research Paris - Aubervilliers (France)
}%

\author{Veneta Grigorova-Moutiers}
\affiliation{
 Saint-Gobain Research Paris - Aubervilliers (France)
}%

\author{Karl Joulain}
\affiliation{
 Institut Pprime, CNRS, Université de Poitiers, ISAE-ENSMA - Poitiers (France)
}%

\date{\today}

\begin{abstract}
Understanding heat transfers in fibrous materials, particularly conduction, is a major challenge due to their heterogeneous and multiscale nature, and the unknown contribution of fiber-to-fiber contacts. In most previous modeling studies, the existence of thermal contact resistance is not considered, and the computational complexity limits the size of simulated samples, which often leads to imprecise or inaccurate predictions. The same problem arises when considering electrical conduction through fibrous materials. In this work, we describe a computationally efficient simulation approach based on multinodal representation to analyze the steady-state heat conduction through the solid structure in numerically generated three-dimensional fibrous networks, including contact resistance. We show that the solid conductivity in these networks is governed by a master curve that depends on a single parameter: a characteristic ratio representing the interplay between the intrinsic fiber conductivity and contact resistance as well as the influence of other geometric parameters, which numerically validates previous theoretical studies. However, we observe a deviation to this established theory for poorly connected networks. We derive an expression for a correction factor, considering the influence of correlations between fiber temperatures, and we then find good agreement with our simulation data. Our results demonstrate that the solid conductivity can be fully predicted based on geometric quantities, regardless of the extent of network connectivity, thus generalizing previous studies on this topic. This work, contributing to improve our understanding of conductive heat transport in fibrous media, may prove useful in the development of accurate predictive models and optimization strategies for fibrous insulation materials.
\end{abstract}

\maketitle


\section{\label{sec:intro}Introduction}

Reducing energy costs in the building sector drives important efforts to improve the performance of insulation materials. These materials are generally porous media, in the form of foams or fibrous assemblies. In particular, glasswool, which is composed of glass fibers dispersed in air and linked together by a binder, is one of the most common insulations currently on the market. The thermal performance of an insulation material can be quantified by its effective thermal conductivity $k_{eff}$. As this quantity refers to a macroscopic scale, it accounts for multiple heat transfer mechanisms: convection, conduction through gas, conduction through the solid phase, and radiation. Improving the efficiency of fibrous insulations requires a quantitative knowledge of these four contributions to heat transport, and more precisely of all the parameters influencing it at the microscale. In media with a pore size below 1~cm, which is the case for glasswool, convection can be neglected~\cite{langlais_isolation_2004}, leaving only the contributions of conduction and radiation, which can be respectively represented by two independent terms $k_{cond}$ and $k_{rad}$. Under the hypothesis of low density, below about 150~kg m$^{-3}$, which is applicable to glasswool, the conductive component can be split as a sum of an air conductivity component $k_{cond,gas}$ and a solid conductivity component $k_{cond,solid}$. This results in the additive approximation~\cite{bankvall_heat_1972} $k_{eff}=k_{rad}+k_{cond,gas}+k_{cond,solid}$, which means that the three identified contributions to heat transfer are assumed to be uncoupled and can be studied separately. Although many studies have been conducted to quantify the radiative contribution in fibrous materials~\cite{tong_analytical_1980,baillis_thermal_2000,arambakam_dual-scale_2013,randrianalisoa_radiative_2017}, only a few have considered the conduction term and its sensitivity to parameter changes. As the contribution of gas conductivity can be considered as constant with the medium’s density and microstructure, it is interesting to further investigate the impact of conduction through the solid phase in fibrous insulation materials, with the aim of understanding and, subsequently, minimizing this contribution. 

\vspace{3mm}

As the separate contributions to the effective conductivity cannot be easily measured experimentally, numerical modeling is the preferred tool for their quantitative analysis. The thermal performance of fibrous materials can be investigated numerically using either real structures reconstructed from images, that can be acquired by X-ray tomography~\cite{meftah_multiscale_2019,faessel_3d_2005,panerai_micro-tomography_2017}, or virtual samples generated by algorithms~\cite{huang_3d_2017,townsend_stochastic_2021,pourdeyhimi_simulation_2006}. The last option enables artificial media to be fitted to real materials using geometric parameters extracted from tomography images, without considering the whole reconstructed three-dimensional (3D) structure for the heat transfer calculations, thus improving computational efficiency.

Several methods have previously been proposed to numerically generate a 3D fibrous network. In some studies, fibers are generated so as to directly avoid each other, i.e., they are considered as hard-core objects, which is the most realistic hypothesis from a physical point of view~\cite{provatas_growth_1997,kulachenko_direct_2012,venkateshan_modeling_2016,naddeo_automatic_2014,redenbach_statistical_2011}. For the sake of simplicity of implementation, soft-core processes, allowing fiber interpenetration, can also be used~\cite{matheron_random_1975,ohser_model-based_2009,gaiselmann_stochastic_2013}. For example, each fiber can be formed by a chain of spheres with a random walk algorithm: this approach, used by Altendorf and Jeulin~\cite{altendorf_random_2010}, can be modified to eliminate fiber overlap~\cite{chapelle_generation_2015,abishek_generation_2017}. An alternative consists in letting the fibers overlap: this implies making no distinction between intersections and contacts, which is a reasonable assumption when the correlations in the spatial distributions of fibers can be neglected~\cite{foygel_theoretical_2005,wang_correlations_2023}. 

The case of fibrous insulation materials such as mineral wools has been previously studied by Chapelle et al.~\cite{chapelle_generation_2015}, Arambakam et al.~\cite{arambakam_simple_2013}, and Kallel and Joulain~\cite{kallel_design_2022}. In order to be as representative as possible in terms of network geometry, the authors chose a given fiber orientation distribution, first introduced by Schladitz et al.~\cite{schladitz_design_2006} for a stacked fiber system. Among its characteristic is an invariance along the $x$ and $y$ axes, representing the morphology of fibrous insulation materials in relation to their manufacturing process. In these studies, a modified random walk generation method was used for each fiber, allowing the successive positions of the spheres to be controlled by two parameters so as to create tortuosity.

\vspace{3mm}

The volume of the 3D fibrous medium that has to be numerically generated should be set to exceed the representative volume element (RVE), defined as the smallest simulation volume for which the computed properties - in this case the solid conductivity - will be representative of the whole material. This solid conductivity can be numerically computed by different methods. Finite-element analysis, used in Refs.~\cite{kallel_design_2022} and~\cite{faessel_3d_2005}, allows a complete representation of the studied structures, resulting in precise yet computationally demanding simulations of the heat transfers, in particular due to the meshing step. An alternative approach to reduce computation time and thus increase the simulation volumes that can be reached is the representation of a fibrous network by a circuit of nodes and branches. This procedure, first used by Kirkpatrick~\cite{kirkpatrick_percolation_1973} to simulate electrical conductivity, is only applicable to the separate calculation of the contribution of conduction through the solid phase, as opposed to volume averaging methods~\cite{whitaker_method_1999}. Although most studies exploiting this method have applied it to quantifying the solid electrical conductivity of metallic nanotube networks~\cite{rocha_ultimate_2015,kim_systematic_2018,kim_analyzing_2021}, very few have extended it to the resolution of thermal transfer problems so far~\cite{ghiaus_causality_2013}.

\vspace{3mm}

Independently of the chosen methods for fiber generation and heat transfer calculation in fibrous media, the dependence of thermal or electrical solid conductivity on network parameters has been numerically investigated in various studies~\cite{kallel_design_2022,kumar_percolating_2005,cheng_transport_2000}, only some of which have included the effect of fiber-to-fiber contact resistance~\cite{dalmas_carbon_2006, zezelj_percolating_2012, behnam_computational_2007, fata_effect_2020}.
It has been found in particular that the Biot number, which compares the conductive thermal resistance in the fiber with the external surface resistance due to thermal losses through contacts~\cite{incropera_fundamentals_2006}, has a noticeable impact on the thermal conductivity of fibrous materials~\cite{vassal_modelling_2008}. 
The reported studies have suggested, in some cases, simplified theoretical models~\cite{zezelj_percolating_2012,kumar_evaluating_2016,forro_predictive_2018,pavlov_conductivity-based_2023}, yet constrained to specific limiting scenarios.
Recently, a statistical analysis combined with Fourier’s law has been used to derive a more general theoretical model that predicts the thermal conductivity of dense networks of straight nanofibers, in two dimensions by Volkov and Zhigilei~\cite{volkov_heat_2012, volkov_thermal_2020}, and in two and three dimensions by Zhao et al.~\cite{zhao_thermal_2018}. This theoretical framework generalizes the results of the abovementioned work by introducing the dependence of solid conductivity on a geometry-dependent resistance ratio. In these studies, the outputs of the model have been exclusively compared with results of numerical simulations in the case of 2D carbon nanotube networks; it is only recently that Volkov and Zhigilei~\cite{ volkov_thermal_2024} presented a first comparison of 3D simulation results, providing at the same time a more accurate theoretical description of semidilute systems. 
However, to the best of our knowledge, there have been no reported studies conducting numerical simulations of 3D networks with variations of all fiber geometric parameters and of contact resistance, making the virtual media representative of a broad spectrum of fibrous materials instead of focusing on an application to metallic nanowires. In particular, anisotropic and poorly connected 3D systems, which represent the typical case of insulation materials, have not been extensively investigated so far, especially theoretically.

\vspace{3mm}

In this work, we focus on quantifying the separate influence of the contribution of conduction through the solid phase to the total heat transfer in fibrous insulation materials, which also includes radiation and conduction through the gas phase. We use a multinodal representation to innovatively study the solid conductivity in virtually generated 3D media with straight fibers, with the specific intention to apply the results to the case of real glasswool samples. A methodological advantage of this method is the possibility of directly integrating a fiber-to-fiber contact resistance as an adjustable parameter, while maintaining reasonable computational efficiency. We demonstrate in particular that the impact of fiber parameters - orientation, length and diameter - as well as that of density, on the conductive properties strongly depends on this contact resistance. More generally, we present an approach to quantify the evolution of the solid conductivity in 3D fibrous networks under the form of a master curve, valid regardless of the individual value of these four fiber parameters. We obtain good agreement between our simulation results and the theoretical model introduced by Zhao et al.~\cite{zhao_thermal_2018} and Volkov and Zhigilei~\cite{volkov_heat_2012,volkov_thermal_2020} for high fiber aspect ratios or high fiber volume fractions, and we show that an adaptation of the model is needed in other cases, in accordance with the latest work from Volkov and Zhigilei~\cite{volkov_thermal_2024}. A first contribution of our work lies in the validation of this modified theoretical model with simulation data, showing its accuracy for the prediction of the solid conductivity for a wide range of 3D networks including anisotropic and poorly connected ones, which goes beyond the commonly studied case of metallic nanowire assemblies. Another innovative discovery is the simple formulation of this model depending on geometrical parameters only, revealing in particular the significance of the average number of contacts per fiber as a key parameter governing the extent of the needed correction compared with the initial high-density case. Finally, we show a comparison between our simulation and theoretical results, focusing on the specific case of glasswool. This approach gives first insights into the potential impact of contacts on thermal performance at the product scale. We also emphasize that our numerical modeling is well suited to predict the electrical conductivity of fibrous media.

\vspace{3mm}

This article is organized as follows. In section \ref{sec:secIIA}, we explain the fibrous medium generation method used in this work. Then, in section \ref{sec:secIIB}, we describe the nodal method used to represent the generated networks by linear circuits and to compute the steady-state heat conduction through the solid phase. Next, numerical results are reported and discussed in section \ref{sec:secIII}; the effect of fiber parameters and contact resistance on the evolution of solid conductivity in 3D fibrous networks is analyzed, and theoretical considerations are introduced based on simulation results. The final section concludes the paper and outlines interesting research directions for future work.

\section{Numerical methods}

\subsection{\label{sec:secIIA}Virtual fibrous medium generation }

\begin{figure*}
\includegraphics[width=1\textwidth]{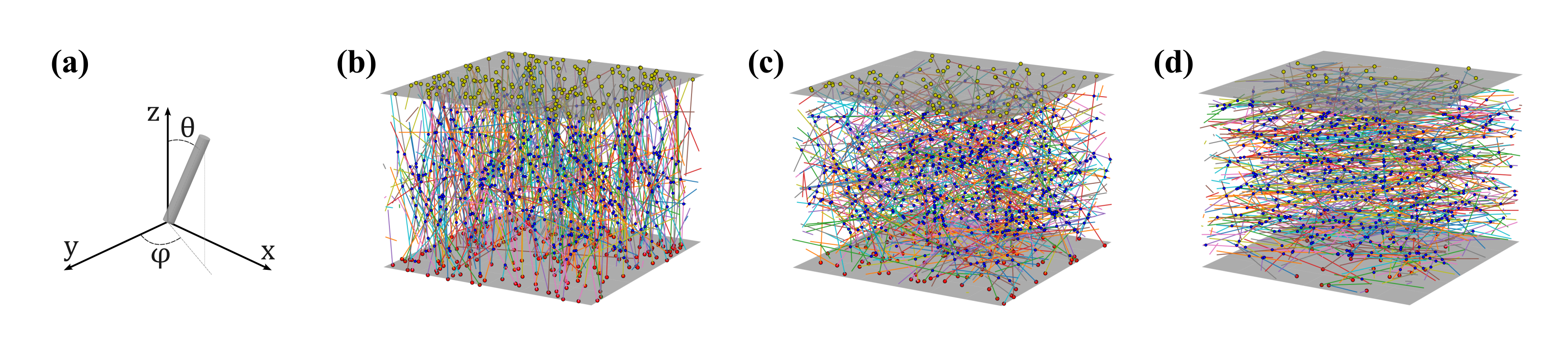}
\caption{\label{fig:fig1} Examples of generated 3D networks, in which the orientation of one fiber $i$ is described by a polar angle $\theta_i$ and an azimuthal angle $\varphi_i$ in the three-dimensional space~(a). In these simulations, $L=3$~mm, the final volume fraction reached is $V_f=0.0033$, the fiber diameter is $d=10~\mu$m, the fiber length is $l=3$~mm and the average polar angle is (b)~$\bar{ \theta}=31.8^\circ$ ($\beta=0.1$), (c) $\bar{ \theta}=57.3^\circ$ ($\beta=1$), and (d) $\bar{ \theta}=79.9^\circ$ ($\beta=8$). The yellow (respectively red) circles represent the contact points between a fiber and the upper (respectively lower) edge of the simulation volume, and the blue circles represent the detected contact points between fibers.}
\end{figure*}

The generated medium is a three-dimensional assembly of straight lines, representing fibers, with an orientation distribution and a prescribed length $l$. This choice is motivated by the low tortuosity of fibers in glasswool as shown by optical coherence tomography acquisitions, an example of which is provided within the Supplemental Material~\cite{supp}. For each line, one extremity is randomly placed in a cubic computational domain of side $L$. Its orientation is defined by a pair of angles ($\theta_i$,$\varphi_i$) in spherical coordinates, as shown in Fig.~\ref{fig:fig1}%
(a), sampled according to a directional distribution, the $\beta$-orientation distribution, previously used in different modeling studies~\cite{schladitz_design_2006,altendorf_random_2010,arambakam_simple_2013,chapelle_generation_2015,kallel_design_2022}.

The azimuthal $\varphi_i$ angle is chosen from a uniform distribution between $0$ and $2\pi$: the density of the distribution is independent of $\varphi$, representing the isotropy of the material in the $x$-$y$ plane.
For the choice of the polar angle $\theta_i$, the following marginal probability density function is applied: 
$g_{\beta} (\theta)=\frac{\beta \sin \theta}{2\lbrack1+(\beta^2-1) \cos^2\beta\rbrack^{3/2}},\theta \in\lbrack0,\pi\rbrack,\beta \in R^{+}\{0\}$. The adjustable parameter $\beta$, called the anisotropy parameter, controls the shape and the mean value of the distribution. The $\beta=1$ case describes an isotropic system; for $\beta\rightarrow0$, the distribution concentrates around the $z$ axis and for $\beta\rightarrow \infty$, the orientations are isotropically distributed in the $x$-$y$ plane. For the sake of convenience, the average polar angle of all fibers $\bar{\theta}$  will be referred to as the network parameter for fiber orientation, rather than $\beta$.

Periodic boundary conditions are applied at the lateral edges of the simulation domain, as their use is known to minimize the RVE size~\cite{altendorf_influence_2014}. When lines, defined by their starting point and orientation, cross a lateral edge of the domain, they are continued on the opposite side until the prescribed length $l$ is reached. However, this condition is not applied to the lower and upper edges, as they will be connected to different heat baths. This means that a small proportion of lines, those in contact with the lower and upper edges, are stopped before reaching the prescribed length, resulting in an average length slightly inferior to $l$. In order to guarantee a small difference between these quantities (less than 15\%), the condition $L$$>$$l$ is applied. It also ensures that no single line is connected to both heat baths at the same time, and that self-interaction between lines due to finite-size effects is limited. In this study, the prescribed length $l$ will be considered as the fiber length, ignoring the small deviation in the mean fiber length that was shown to have a negligible impact on the final results.

The number of lines initially generated in the system is chosen according to a preimposed volume fraction, called the initial volume fraction, and considering them as cylinders with a fixed value $d$ for their diameter. The contacts are then identified using a distance threshold: two lines are considered connected if the smallest distance between them is inferior to their diameter $d$. In this approximation, the connected fibers are modeled by soft-core, i.e., interpenetrating, objects. The soft-core approach does not consider correlations in the spatial distribution of lines, as they may exist in a physical system of connected fibers. However, it has been reported in previous studies~\cite{foygel_theoretical_2005,berhan_modeling_2007} that the error introduced by this approximation is small, less than 1\% for an aspect ratio $a=l/d\geq100$, which is the average one in glasswool~\cite{noauthor_private_2024}. It should be noted that lines having less than two contact points cannot be included in the nodal analysis described in the following, as the resulting thermal network has to be fully connected to be solved by the matrix approach. As they do not contribute to heat transfer, after contact detection, these lines are removed from the generated medium, as well as the associated contact points through an iterative procedure. In this study, the volume fraction of medium $V_f$ is defined as the final one, obtained after fiber removal and therefore inferior to the initial one. Figures~\ref{fig:fig1}(b), \ref{fig:fig1}(c) and \ref{fig:fig1}(d) show generated networks using our numerical approach with identification of contact points, for three different angle distributions controlled by parameter $\beta$.

\subsection{\label{sec:secIIB}Nodal analysis for the resolution of a steady-state conductive heat transport problem}

The steady-state conduction through the solid phase in the generated network is computed numerically using the configuration of the guarded hot plate experiment. The upper and lower boundaries of the simulation domain represent two plates with different temperatures, which means that each fiber extremity located at $z=0$ (respectively $z=L$) will be a temperature source at $T=T_b$ (respectively $T=T_b  + \Delta T$), while the fibers will be considered as conductive with an intrinsic thermal conductivity $k_{fib}=1.3$~Wm$^{-1}$K$^{-1}$. This value corresponds to the thermal conductivity of silica glass, the fiber material in glasswool mats. The four other boundaries, on the side of the simulation domain, are assumed to be adiabatic. In the isotropic case, the solid thermal conductivity is the same in each direction. However, if the anisotropy parameter of the distribution orientation is different from unity, the transverse component (i.e., the thermal conductivity in the $z$ direction) differs from the planar component (i.e., that in the $x$ or $y$ direction). This is particularly the case for fibrous insulation materials, due to their layered fiber arrangement~\cite{akolkar_tomography_2017}. Note that here, in the guarded hot plate experiment configuration, we exclusively simulate the transverse solid thermal conductivity, referred to in this work as the solid conductivity, meaning that the heat flux direction will always be the cross-plane direction for the fibrous medium.

\begin{figure}
\includegraphics[width=0.48\textwidth]{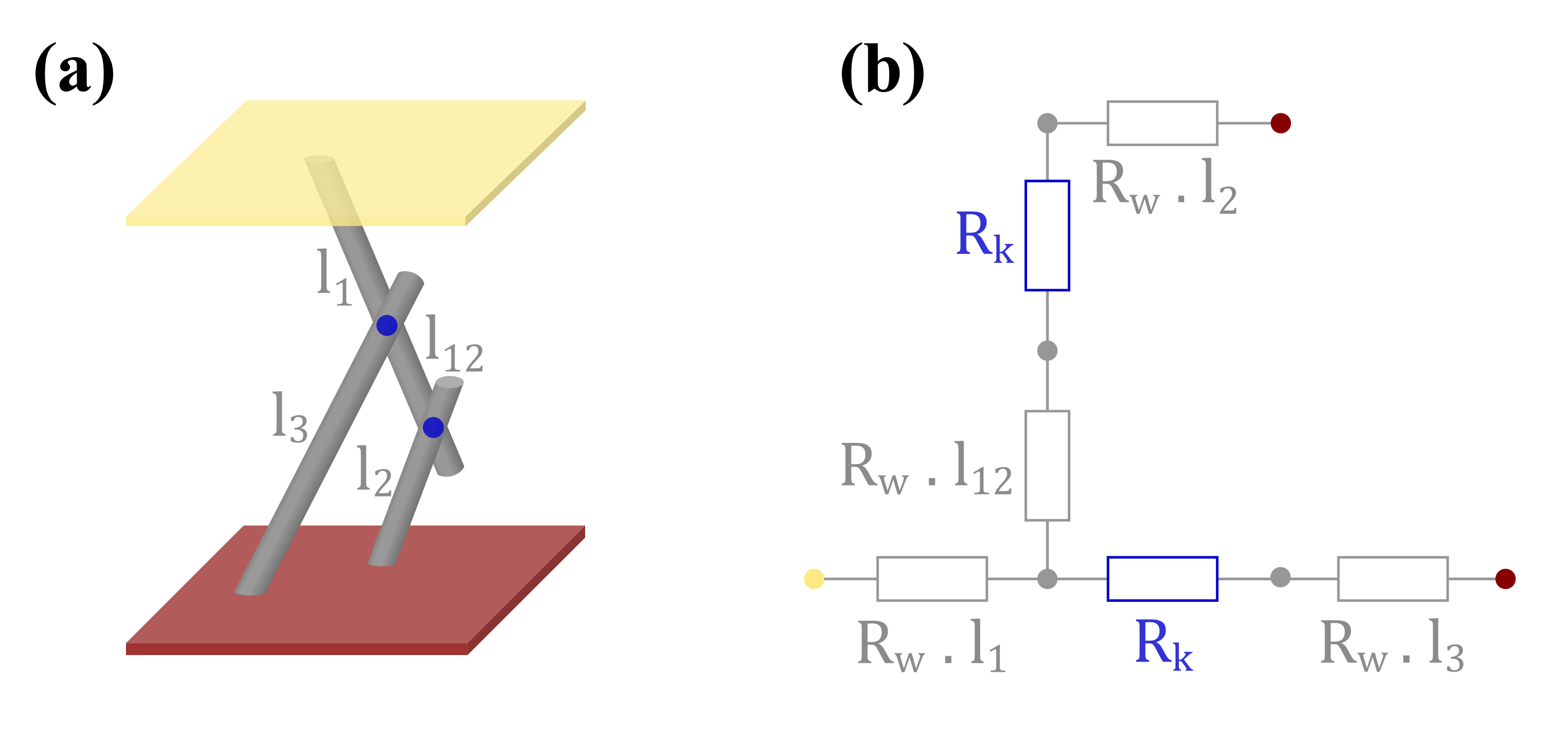}
\caption{\label{fig:fig2} (a)~Simplified network of three connected fibers. Contacts are identified with blue circles. (b)~Equivalent thermal circuit of resistances in the steady-state configuration. The dots represent nodes; in particular, yellow and red dots correspond to temperature sources. A given branch $f$ can be associated with either a contact resistance $R_k$ or an internal resistance $R_w l_f$ proportional to its length, where $R_{w}= \frac {4}{k_{fib} \pi d^2 }$.}
\end{figure}

The generated network has to be sufficiently dense or large to reach the RVE and to allow for a low-incertitude prediction of the solid conductivity. By analogy with an electrical circuit, it can be computed using graph theory, with the modified nodal analysis method~\cite{kim_systematic_2018}. In the steady state, this thermal network is a linear circuit arranged in branches, representing heat flow rates crossing thermal resistances, connected to each other by nodes representing temperatures or temperature sources~\cite{ghiaus_causality_2013,kaveh_introduction_2013}. This nodal method assumes uniform sections in temperature, i.e., the conductive heat transfer in each fiber will be considered as unidimensional, occurring only in the direction of the fiber. This choice is justified by the average aspect ratio of fibers $a\approx100$ in glasswool~\cite{noauthor_private_2024}. In particular, the network is modeled using a multinodal representation, first introduced by Da Rocha et al.~\cite{rocha_ultimate_2015} for electrical circuits, considering both the internal thermal resistance of fibers and the thermal contact resistance. This modeling method is represented in Fig.~\ref{fig:fig2}%
.
In this kind of model, each node represents a contact point: either between a fiber and a temperature source on the upper or lower edge of the simulation cube, or between two fibers. In the latter case, the junction is defined by two distinct nodes linked by an additional branch that represents the fiber-to-fiber contact resistance. This kind of model differs from the junction resistance-dominated assumption, in which one node represents one fiber, linked by contact resistances, thus ignoring the contribution of the internal resistances. Also, as can be seen in Fig.~\ref{fig:fig2}, when fiber extremities are not touching any other fiber, they are assumed to be adiabatic and are therefore not included in the thermal network.

\begin{figure*}
\includegraphics[width=0.9\textwidth]{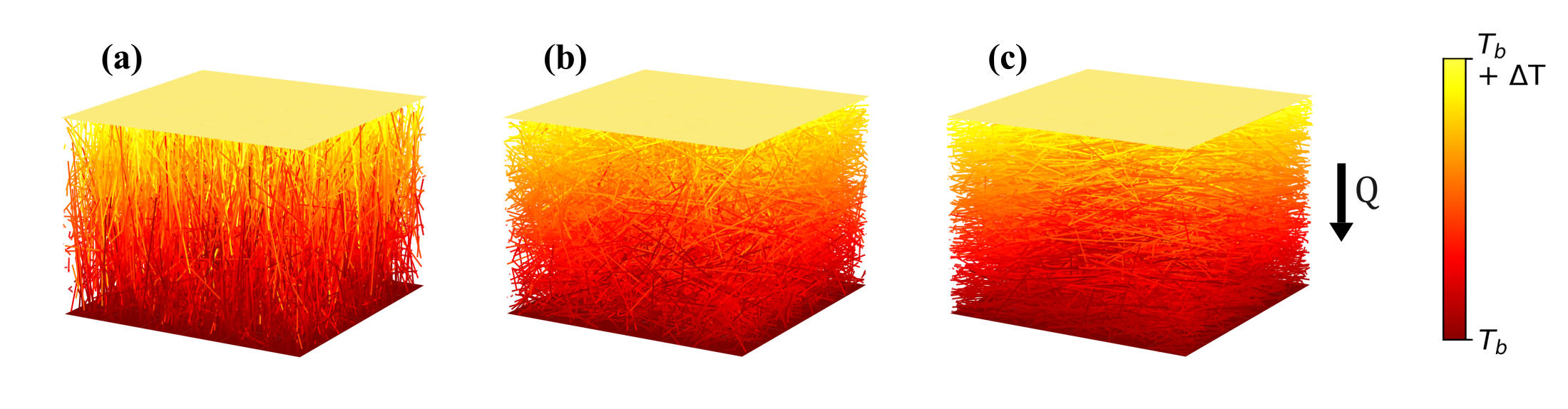}
\caption{\label{fig:fig3} Temperature field distributions in generated networks with $L=3$~mm, $V_f=0.0198$, $l=2$~mm, $d=10~\mu$m, and (a) $\bar{ \theta}=31.8^\circ$ ($\beta=0.1$), (b) $\bar{ \theta}=57.3^\circ$ ($\beta=1$), and (c) $\bar{ \theta}=79.9^\circ$ ($\beta=8$), containing between 3720 and 3990 fibers. The upper and lower edges of the simulation domain are areas where the temperature is imposed, generating a heat flux $Q$ in the vertical direction. The angle $\theta_i$ of each fiber $i$ is therefore its angle with respect to the direction of the heat flux. The coordinate system used is that represented in Fig.~\ref{fig:fig1}(a).}
\end{figure*}

As the considered heat transfer problem is a steady-state configuration, the associated thermal network can be described by algebraic graph theory. The resolution of a single matrix equation allows us to get the vector of temperatures and that of heat fluxes, making possible the representation of the temperature field in the network, as shown in Fig.~\ref{fig:fig3}%
 for different angle distributions (see Appendix~\ref{sec:appA} for details on the resolution method). Then, the total heat flux $Q$ is computed by adding the contributions from the branches in contact with the upper or lower edge of the simulation domain, and the solid thermal conductivity is calculated as $k_{cond,solid}=QL/(L^2\Delta T)$. As the thermal properties of individual fibers, the fiber conductivity $k_{fib}$ and the fiber-to-fiber contact resistance $R_k$, are assumed to be independent of the temperature, the solid conductivity does not depend on $T_b$ and $\Delta T$. 

In order to lower the uncertainty of the simulation results, the selected value of the solid thermal conductivity for a defined set of parameters is the mean of $k_{cond,solid}$ over five realizations; its associated 95\% confidence interval is also computed and is represented under the form of error bars in all figures shown in this paper. This confidence level is a common choice in statistics: it has been used in particular in previous studies on similar fibrous systems, with $n\approx10$ realizations~\cite{altendorf_influence_2014,chapelle_phd_2016}. In this work, errors are calculated using the Student's probability distribution, taking into account the fact that the standard deviations of samples of $n=5$ realizations underestimate the true Gaussian standard deviations. The relative error $\epsilon_{rel}$ on $k_{cond,solid}$ for each simulation point is obtained from the expression $\epsilon_{rel}=\frac{t_{n-1}(1-\alpha/2) s}{\overline{k_{cond,solid}} \sqrt{n}}$ where $\alpha$ is the significance level; here $\alpha=0.05$ to achieve a 95\% confidence interval. In this expression, $\overline{k_{cond,solid}}$ is the sample mean, $s$ is the unbiased estimator of the standard deviation given by~$s=\sqrt{\frac{\sum_{i=1}^{n} (k_{cond,solid}^i-\overline{k_{cond,solid}})^2}{n-1}}$, and $t_{n-1}(1-\alpha/2) = 2.78$ is the critical value of the $t$ distribution for a 95\% confidence level. Here $n-1$, the number of realizations minus one, represents the number of degrees of freedom, i.e, the number of values that are allowed to vary when estimating the statistic.

For each set of parameters, the sample volume $V=L^3$ is chosen so as to exceed the RVE, on top of the previous condition $L> l$. In this study, the RVE is defined as the volume for which the relative error on the solid thermal conductivity $\epsilon_{rel}$ introduced above is smaller than $5\%$ for $R_k=0$. This relative error criterion has been selected as a similar approach was used by Altendorf et al.~\cite{altendorf_influence_2014} for the selection of RVE in fibrous media; it also ensures a reasonable computation time. The procedure for selecting $L$ is illustrated in Fig. S2 within the Supplemental Material~\cite{supp}: starting from a domain size slightly larger than the fiber length, it is gradually increased until the average of $k_{cond,solid}$ for $R_k=0$ on five realizations satisfies the error criteria $\epsilon_{rel}<5\%$, which also corresponds to achieving convergence of the results. This ensures that the simulation results represent the whole network and are not dominated by the behavior of individual fibers.  As a consequence of this procedure, the chosen value of $L$ depends on the input parameters; this is represented in Fig. S3 within the Supplemental Material~\cite{supp}. To reach the RVE, a larger domain size is required when the volume fraction decreases, or when fiber dimensions increase, all other parameters being fixed, which can be attributed to the decrease of the total number of fibers. In contrast, the average fiber angle appears to have only a minor impact on the selection of $L$.

Numerical simulations are performed to independently study the influence of geometric fiber parameters on the resulting thermal properties of the network. In particular, the joint effects of the fiber volume fraction $V_f$, fiber length $l$, fiber diameter $d$, and fiber orientation $\bar{ \theta}$ with contact resistance $R_k$ are investigated. For each set of parameters and on top of the conditions on $V$ and on $L$ described previously, as some fibers are deleted in the generation process, the orientation and length distributions are compared between the initial and final states, to ensure that they are not substantially modified. For all simulations in this study, the size of the calculation domain ranges from $L=1.25$~mm to $L=6$~mm depending on the four parameters, the contact density ranges from 40 to 20000~mm$^{-3}$, the average number of contacts per fiber ranges from 2 to 20, and the fiber density ranges from 25 to 4000~mm$^{-3}$.

\section{\label{sec:secIII}Results and discussion}

\subsection{\label{sec:secIIIA}Effect of thermal contact resistance on the solid conductivity in relation to fiber properties}

\subsubsection{\label{sec:secIIIA1}Effect of thermal contact resistance at different fiber volume fractions}

\begin{figure}[b]
\includegraphics[width=0.48\textwidth]{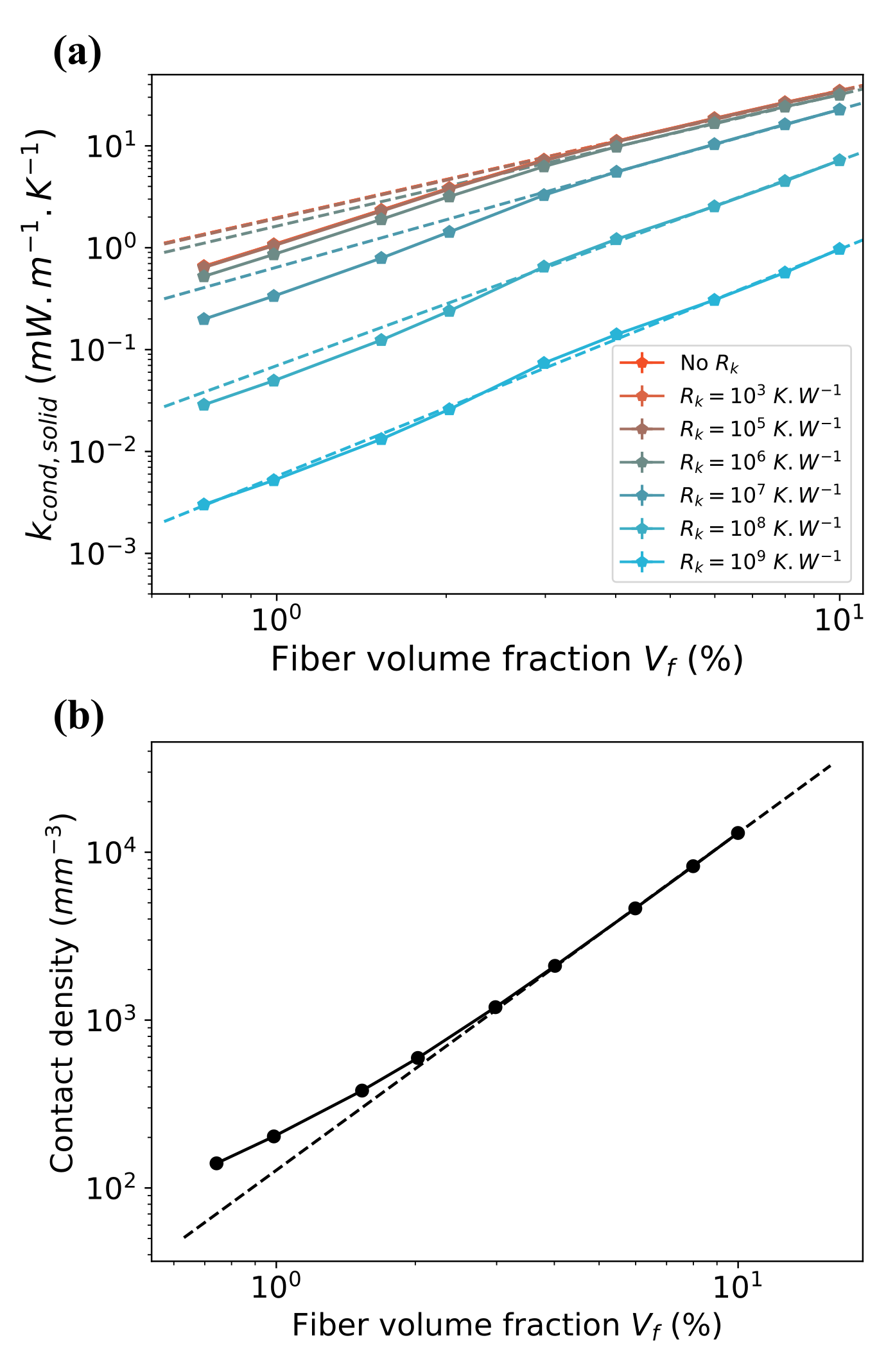}
\caption{\label{fig:fig4} (a)~Solid conductivity as a function of the fiber volume fraction for different thermal contact resistances. The dashed lines indicate the power-law best fits to the simulation data for $0.03\geq V_f\geq 0.1$, with exponents ranging from 1.25 to 2.23 as the contact resistance increases from $0$ to $10^9$~K~W$^{-1}$. The constant input parameters are $l=1$~mm, $d=10~\mu$m, and $\bar {\theta} =57.3^\circ$. (b)~Plot of the corresponding contact densities, as a function of fiber volume fraction. The dashed line indicates the square power-law best fit to the simulation data for $0.03\geq V_f\geq 0.1$.}
\end{figure}

The variation of the solid conductivity with the fiber volume fraction $V_f$, fixing the fiber orientation, length, and diameter, is illustrated in Fig.~\ref{fig:fig4}%
(a)
for different contact resistances. Volume fractions between 0.0075 and 0.1 are studied, corresponding to glasswool mat densities ranging from 17 to 221~kg~m$^{-3}$. In all cases, the solid conductivity increases with density in a nonlinear way; in particular, it shows power-law behavior at high volume fractions ($V_f\geq0.03$) indicated by the dashed line best fits. The graph also shows that the increase in the solid conductivity with the volume fraction becomes more pronounced as the contact resistance increases, i.e., the exponent in the power law increases, from 1.25 for $R_k=0$ to 2.23 for $10^9$~K~W$^{-1}$. This generalizes previous results reported in Refs.~\cite{zezelj_percolating_2012} and~\cite{fata_effect_2020} for 2D networks to the case of 3D networks. These studies have also distinguished two regimes depending on volume fraction: the low-density regime, close to the percolation threshold, exhibits a different behavior from that observed at higher densities. This is presumably due to a difference in the degree of connectivity of the networks, i.e., in the proportion of fibers and contacts conveying heat. Indeed, the contact density increases when the volume fraction increases, as shown in Fig.~\ref{fig:fig4}(b). For high densities, it is expected to follow a square power law~\cite{zezelj_percolating_2012}, which is verified in our simulation results for $V_f\geq0.03$. This value of the volume fraction provides a first estimate of the high-density threshold for the chosen simulation parameters, consistent with the validity range of the power laws introduced in Fig.~\ref{fig:fig4}(a). According to this estimation, networks with $V_f=0.01$ should correspond to a low-density regime in which the connectivity is low: to simplify the study of the impact of the other fiber parameters, this value of the volume fraction will be chosen in the following sections.

\subsubsection{\label{sec:secIIIA2}Effect of thermal contact resistance at different fiber orientations}

\begin{figure}[b]
\includegraphics[width=0.48\textwidth]{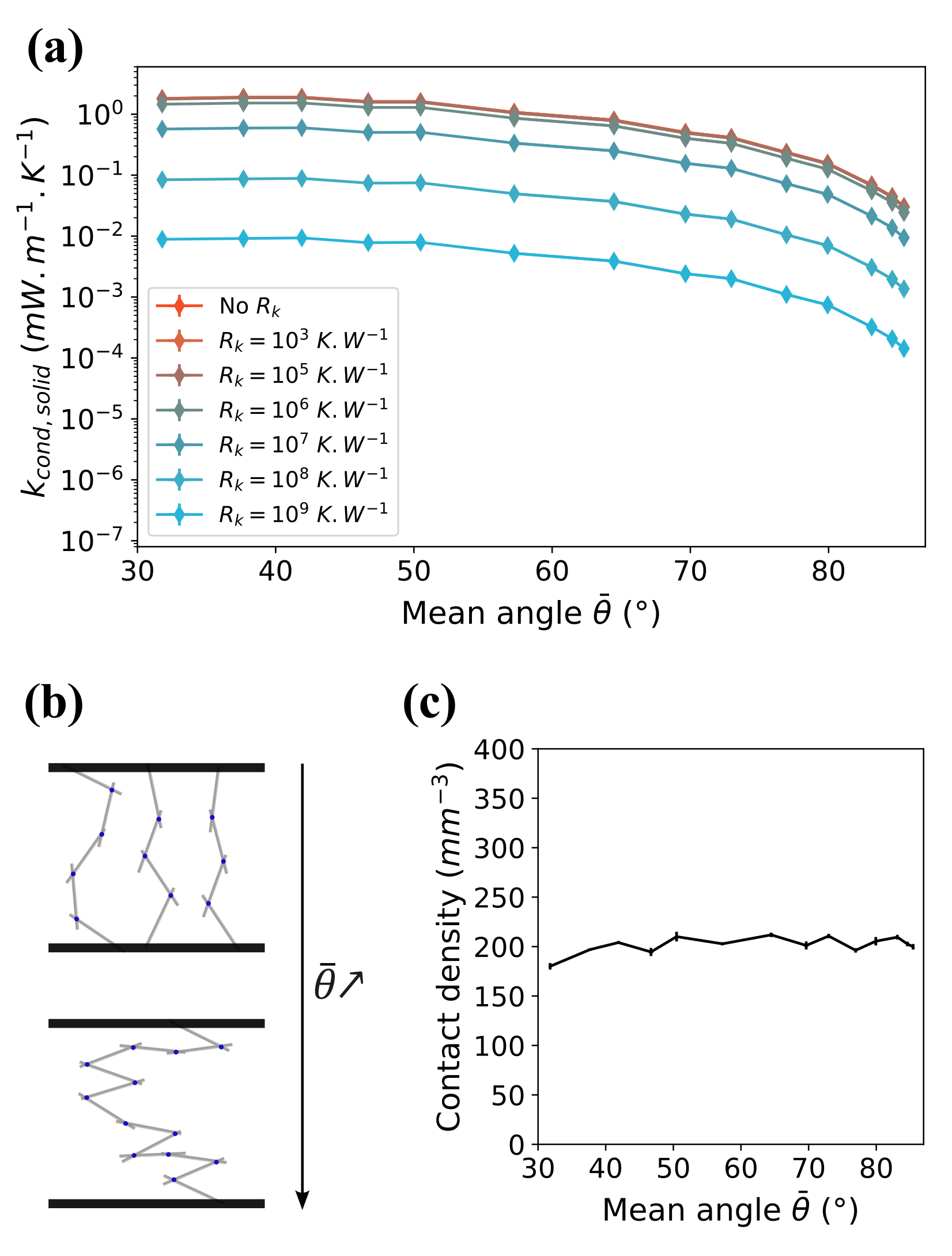}
\caption{\label{fig:fig5} (a)~Solid conductivity as a function of the mean angle to heat flux direction for different thermal contact resistances. The anisotropy parameter $\beta$ ranges from 0.1 to 30, corresponding to mean angles $\bar{\theta}$ between 31.8$^\circ$ and 85.5$^\circ$: only data obtained for $\beta>1$, i.e., for networks in which fibers are preferentially oriented in the cross-plane direction, are representative of insulation materials such as glasswool. The constant input parameters are $l=1$~mm, $d=10~\mu$m, and  $V_f=0.01$. (b)~Representation of simplified networks illustrating a qualitative argument for the dependence of the conductivity on the mean angle to heat flux direction: when it increases, the number of conduction paths decreases. (c)~Plot of contact densities, corresponding to the simulations of (a), as a function of the mean angle. }
\end{figure}

Figure~\ref{fig:fig5}%
(a)
illustrates the variation of the solid conductivity as a function of the mean angle between the fibers and the heat flux direction $\bar{\theta}$, for different values of the contact resistance and for fixed fiber length, diameter, and volume fraction. The mean angle $\bar{\theta}=57.3^\circ$ corresponds to a uniform distribution in spherical coordinates, i.e., to $\beta=1$; then $\beta$ is increased up to 30, corresponding to a mean angle of 85.5$^\circ$, so as to cover the estimated range for $\bar{\theta}$ in glasswool, between 75$^\circ$ and 85$^\circ$~\cite{noauthor_private_2024}. For the sake of completeness, we have also included in this figure the simulation results obtained for $\beta<1$. As $\beta$  decreases below 1 and approaches 0, fibers become increasingly aligned approximately parallel to the $z$~axis, as can be seen in Figs.~\ref{fig:fig1} and~\ref{fig:fig3}. Note that this case is not representative of insulation materials, in which fibers tend to form significant angles to the heat flux direction. The graph shows that the solid conductivity decreases as the mean angle to the heat flux direction increases whether contact resistances are integrated or not. This indicates that the impact of contacts on the solid conductivity does not change with the orientation of fibers, including when they tend to align with the heat propagation axis.

Note that at sufficiently low densities, the interconnected conduction paths are few in number and can be neglected as a first approximation. In this case, a simplified expression for the total thermal resistance can be built based on simplified networks such as those represented in Fig.~\ref{fig:fig5}(b), accounting for the contributions of several conduction paths in parallel, each consisting of several fibers in series. The total resistance $R_t$ of such networks is the ratio of two parameters: the average resistance of the paths, $R_p$, and the number of conductive paths in parallel, $n_p$~\cite{park_modeling_2019,kumar_current_2017}. This average path resistance $R_p$ is the result of two factors that add up as a series: the total resistance of contacts, and the total internal resistance of the path. Therefore,
\begin{eqnarray}
R_t=\frac{1}{n_p}\left(n_{c,p}R_k  + \frac {4 l_p}{\pi k_{fib}d^2 }\right)
\label{eq:one}
\end{eqnarray}
or
\begin{eqnarray}
R_t=\frac{1}{n_p^2}\left(n_{c,tot}R_k  + \frac {4 l_{tot}}{\pi k_{fib}d^2 }\right)
\label{eq:two}
\end{eqnarray}
where $n_{c,p}$ is the average number of contacts per path, $n_{c,tot}$ is the total number of contacts, $l_p$ is the average length of a path and $l_{tot}$ is the total length of the paths. In these considerations, the unconnected fiber ends are neglected: therefore, it is assumed that $l_{tot}$ is the total length of fibers in the networks. As shown in Fig.~\ref{fig:fig5}(b), more fibers are needed to connect one heat bath to the other through one path when $\bar{\theta}$ increases. Therefore, the created paths are overall less direct and longer, leading to a decrease in the total number of paths $n_p$. As the volume fraction and the fiber dimensions are fixed in the simulations, the total fiber length is constant: Eq.~(\ref{eq:two}) confirms that $R_t$ increases with $\bar{\theta}$ when there is no contact resistance, leading to a decrease in the conductivity. Moreover, the parameter governing the relative impact of contacts with respect to that of internal resistance is the total number of contacts $n_{c,tot}$. Figure~\ref{fig:fig5}(c) shows that the volume density of contacts remains around 200~mm$^{-3}$ regardless of fiber orientation for the simulation parameters chosen in this study: this supports our observation that the effects of contacts and of the fiber mean angle on the network are not correlated in the studied range, and particularly in that representative of glasswool.

\subsubsection{\label{sec:secIIIA3}Effect of thermal contact resistance at different fiber lengths}

\begin{figure}[b]
\includegraphics[width=0.48\textwidth]{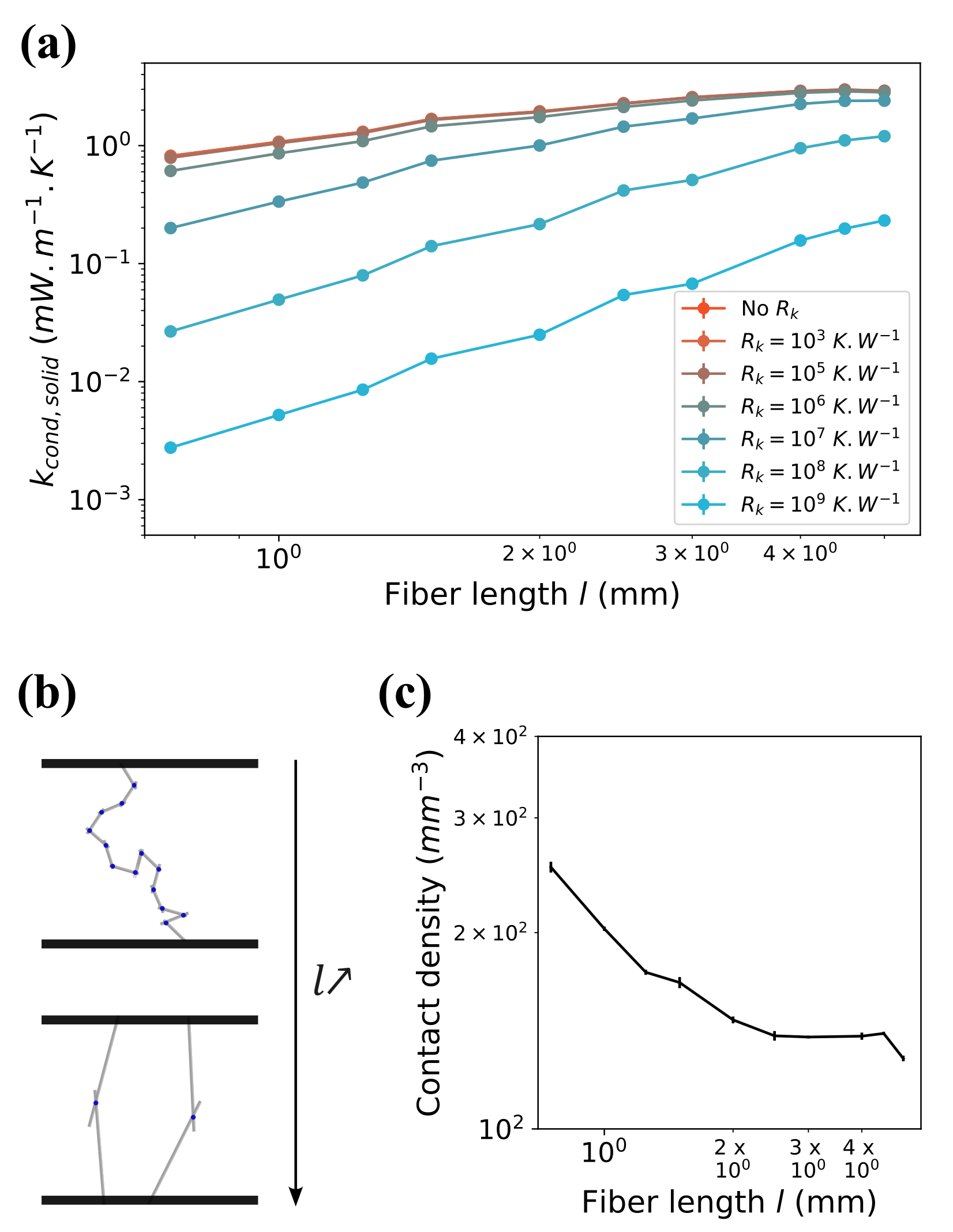}
\caption{\label{fig:fig6} (a)~Solid conductivity as a function of fiber length for different thermal contact resistances. The constant input parameters are $\bar {\theta} =57.3^\circ$, $d=10~\mu$m, and  $V_f=0.01$. (b)~Representation of simplified networks illustrating a qualitative argument for the dependence of the conductivity on fiber length: when it increases, the number of conduction paths increases and the number of contacts per path decreases~\cite{hecht_conductivity_2006}. (c)~Plot of contact densities, corresponding to the simulations of (a), as a function of fiber length.}
\end{figure}

Figure~\ref{fig:fig6}%
(a)
displays a plot of the solid conductivity versus the fiber length $l$ for different contact resistances, fixing the other fiber parameters. The fiber length ranges from 0.75 to 5~mm in the simulations, remaining within the same order of magnitude of 1~mm, which is representative of glasswool~\cite{noauthor_private_2024}. The graph shows that, when no contact resistance is considered, the solid conductivity increases when the fiber length increases. Including the impact of junctions between fibers, our results show that an increasing contact resistance induces a decrease in the solid conductivity compared with its value at $R_k=0$, and that the impact of fiber length is amplified when the contact resistance increases. This means that contact resistance has a lower relative impact for longer fiber assemblies. This behavior is in agreement with the results obtained in Ref.~\cite{fata_effect_2020} for electrically conducting 2D nanowire networks.

The density and diameter being fixed, the total length is constant and the number of fibers needed to create a path decreases when $l$ increases. Conduction paths are becoming less tortuous and shorter: as a consequence, more conduction paths in parallel can convey heat. According to Eq.~(\ref{eq:two}), this explains the observed increase in the solid conductivity when there is no contact resistance; Fig.~\ref{fig:fig6}(b) illustrates this situation. Equation~(\ref{eq:two}) also shows that when it comes to the impact of contacts in simplified networks, the total number of contacts is the parameter that determines its relative importance with respect to the impact of internal resistance, similar to the case of fiber orientation variation. The evolution of contact density versus fiber length is presented in Fig.~\ref{fig:fig6}(c). For $l<2.5$~mm, contact density decreases with fiber length. Then, it seems to reach a threshold - the onset of decline observed at $l>4.5$~mm remains to be confirmed by further simulations. This behavior is in accordance with previously established theoretical models that estimate the number of contacts in a network of lines~\cite{komori_numbers_1977}. Overall, we find that contact density tends to decrease over the studied range. According to Eq.~(\ref{eq:two}) and as represented in the examples of Fig.~\ref{fig:fig6}(b), this should lead to a reduction of the resistive contribution of contacts in the network as there are fewer contacts per path, which is consistent with the correlation observed in Fig.~\ref{fig:fig6}(a) between the impact of contacts and the impact of fiber length.

\subsubsection{\label{sec:secIIIA4}Effect of thermal contact resistance at different fiber diameters}

\begin{figure}[b]
\includegraphics[width=0.48\textwidth]{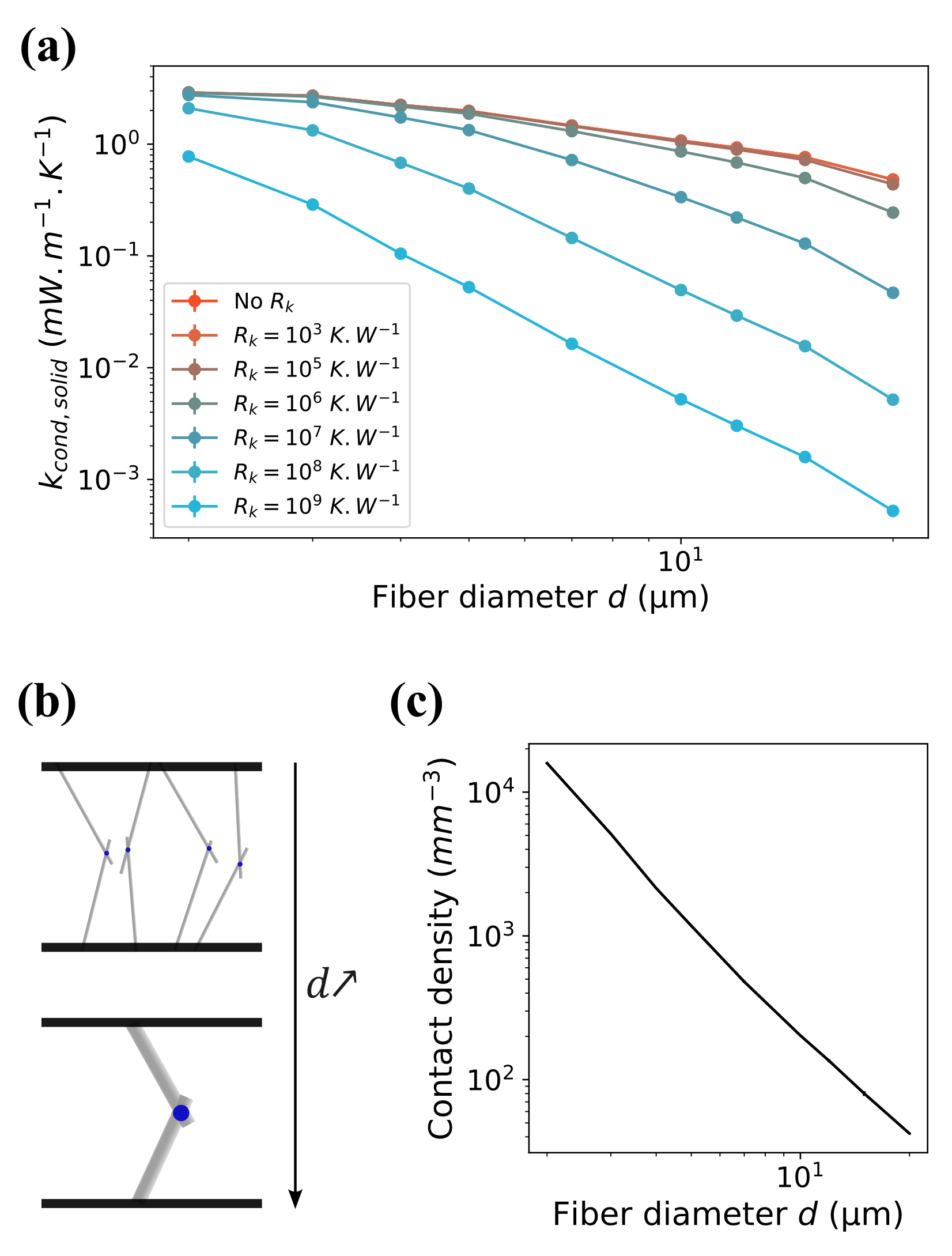}
\caption{\label{fig:fig7} (a)~Solid conductivity as a function of fiber diameter for different thermal contact resistances. The constant input parameters are $\bar {\theta} =57.3^\circ$, $l=1$~mm, and  $V_f=0.01$. (b)~Representation of simplified networks illustrating a qualitative argument for the dependence of conductivity on fiber diameter: when it increases, the number of conduction paths in parallel decreases. Assuming a constant number of contacts per path, the number of contact resistances $R_k$ in parallel decreases, leading to an increase in the total resistive contribution of junctions, which goes from $R_k/4$ to $R_k$ in this example~\cite{hecht_conductivity_2006}. (c)~Plot of contact densities, corresponding to the simulations of (a), as a function of fiber diameter. }
\end{figure}

Note that even though widthless sticks are considered in the generation method, a diameter $d$ is taken into account both through the contact detection threshold and the expression of the intrinsic resistance of fibers $R_{fib}= \frac {4l}{k_{fib} \pi d^2 }$. Its influence on the solid conductivity for different contact resistances is represented in Fig.~\ref{fig:fig7}%
(a).
The diameter ranges from 2 to 20~$\mu$m so as to cover the sizes of the glass fibers that can be found in real glasswool mats~\cite{noauthor_private_2024}. The graph shows that, with no contact resistance, the solid conductivity decreases when the fiber diameter increases. Also, similar to the case of fiber length, the impact of fiber diameter is amplified when the contact resistance increases.

Our numerical results indicate that among the four ranges of variation for fiber parameters studied individually, the one chosen for the fiber diameter leads to the greatest variation in the total number of fibers. This means that changing the diameter between 2 and 20~$\mu$m greatly affects the connectivity of the network. As a consequence, the assumption of low density in the sense of no interconnection between conduction paths cannot be made here over the whole diameter range, even though the volume fraction is fixed at 0.01. Networks with a higher diameter are composed of less fibers, and therefore exhibit less connectivity, leading to a less efficient heat flux distribution which explains the observed decrease in the solid conductivity. This effect is expected to be important as the fixed volume fraction is low~\cite{bergin_effect_2012}.  Moreover, once again Eq.~(\ref{eq:one}) or~(\ref{eq:two}) gives insights into the evolution of the relative impact of contacts with respect to that of the internal resistance with fiber diameter, which is illustrated by simplified networks in Fig.~\ref{fig:fig7}(b). The total contact density decreases as the fiber diameter increases – this is verified by simulation results shown in Fig.~\ref{fig:fig7}(c), and the number of conduction paths in parallel decreases as well. From Eq.~(\ref{eq:one}), assuming constant $l_p$ and $n_{c,p}$, an increase in the fiber diameter increases the ratio of the resistive contribution of contacts over that of paths due to its dependence on $d^2$. This leads to an amplification of the impact of the remaining contact resistances, even though there are fewer of them. These considerations, although they are limited to a simplified case with no interconnection, are consistent with the correlation between the impact of contacts and impact of fiber diameter observed in Fig.~\ref{fig:fig7}(a). 

\vspace{3mm}

Note that while in this study contact resistance and fiber diameter are treated as independent, it is likely that a correlation exists between these two parameters~\cite{hecht_conductivity_2006}. Indeed, glass fibers are expected to deform at a junction, according to the Hertz contact law. Therefore, the contact area between fibers with a larger diameter would increase, leading to a lower associated contact resistance. In the configuration studied here, which can be described as a perfectly rigid assumption, one can note that the effects of fiber length and diameter seem to have inverse effects, indicating a potential dependence of the network properties on the aspect ratio $a$. In particular, the influence of the aspect ratio on the solid conductivity is amplified in the case of resistive contacts, which was also reported in Ref.~\cite{vassal_modelling_2008}.

\subsection{\label{sec:secIIIB}Unified description of the evolution of the solid conductivity }

As the conductivity of fibers $k_{fib}$ is fixed in the simulations discussed in the previous section, changing the value of the contact resistance $R_k$ amounts to changing the resistance ratio $R_k/R_{fib}$ where $R_{fib}=\frac {4l}{k_{fib} \pi d^2}$ is the intrinsic resistance of fibers.  We verified with additional simulations that modifying the value of $R_k/R_{fib}$ by increasing $R_k$ or decreasing $k_{fib}$ gives the same results, for an identical network structure. Besides the resistance ratio, our results demonstrate that fiber parameters also participate in the control of conduction in the network. A relevant measure of this interplay between the intrinsic and contact resistances that takes into account the network geometry is given by a dimensionless parameter $r$ introduced in a theoretical model by Zhao et al.~\cite{zhao_thermal_2018} and Volkov and Zhigilei~\cite{volkov_heat_2012,volkov_thermal_2024}:

\begin{eqnarray}
r=\frac {2l\langle\lvert \cos\theta\rvert \rangle}{\langle H \rangle \langle N_c \rangle} \frac {R_k}{R_{fib}} = \frac {R_k \langle\lvert \cos \theta\rvert \rangle k_{fib} \pi d^2} {2\langle H \rangle\langle N_c \rangle}
\label{eq:three}
\end{eqnarray}.
Here $\langle N_c \rangle$ is the average number of contacts per fiber, $\langle H \rangle$ is the average vertical distance between two centers of fibers in contact, and $\langle\lvert \cos \theta\rvert\rangle$ is the average cosine of the polar angle of fibers. As highlighted by the authors, the theoretical models for 2D and 3D fibrous networks are similar, as a 3D network becomes 2D if the azimuthal angle reaches $\varphi=\pi/2$. Note that this transition from 3D to 2D networks is not investigated in this study as the azimuthal angle is always chosen from a uniform distribution: the distribution of polar angles $\theta$ is the only one that can change. According to this model, the solid conductivity in a 3D fibrous network is given by

\begin{equation}
k_{cond,solid}=\frac{1}{4} k_{fib} \pi d^2 \langle n_z \rangle \langle\lvert \cos \theta\rvert\rangle \frac{1}{1+r}=k_{0,th} \frac{1}{1+r},
\label{eq:four}
\end{equation}
\vspace{0.2mm}

\noindent where $\langle n_z \rangle$ is the average areal number density of fibers through a horizontal section, and $k_{0,th}=\frac{1}{4} k_{fib} \pi d^2 \langle n_z \rangle \langle\lvert\cos \theta\rvert\rangle$ is the theoretical value of the conductivity at $R_k=0$. This model is based on the assumptions of linearity of the temperature distributions along fibers, and of noncorrelations between the temperatures of two fibers in contact.

\vspace{3mm}

\begin{figure}[b]
\includegraphics[width=0.48\textwidth]{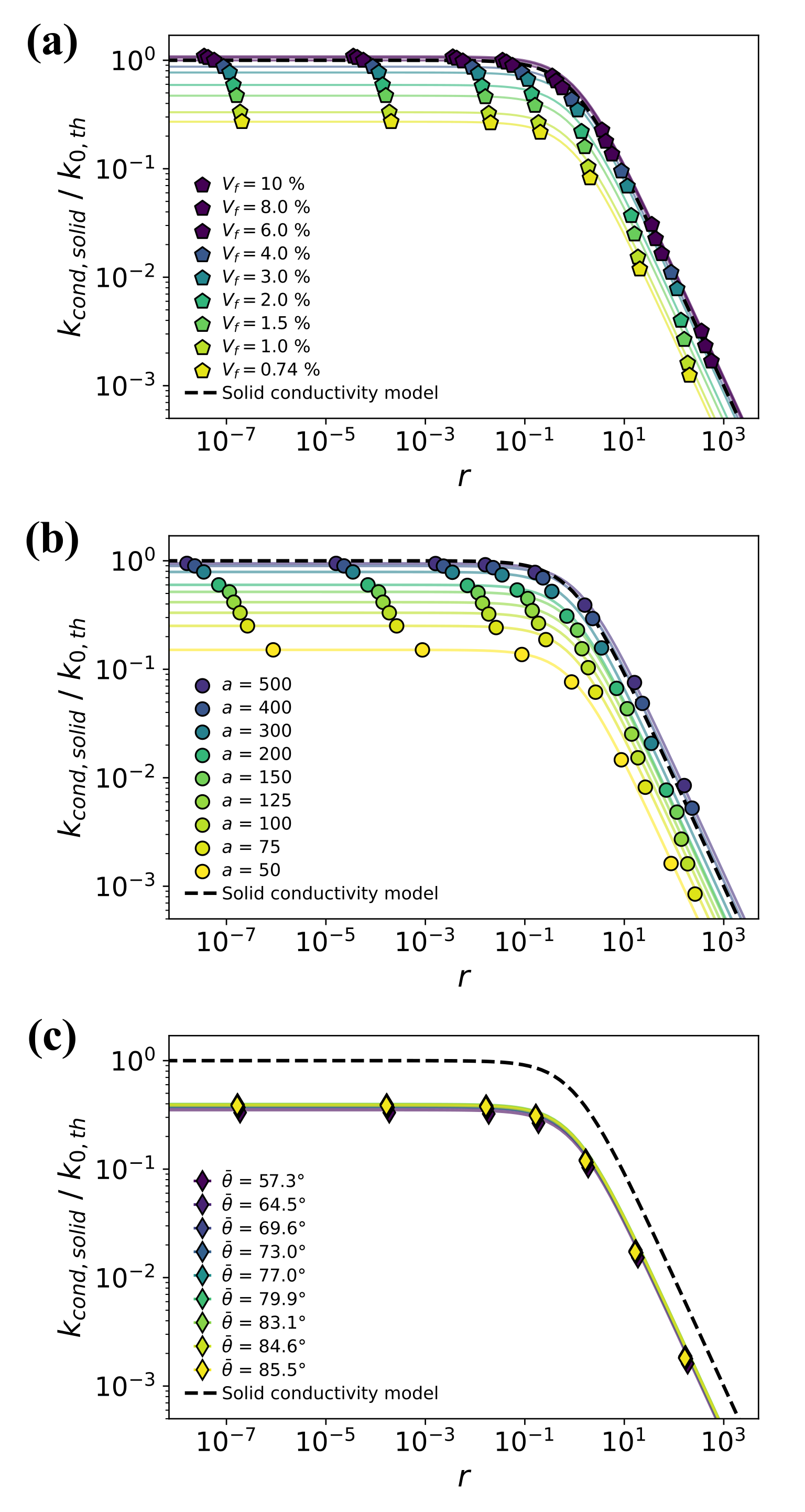}
\caption{\label{fig:fig8} Solid thermal conductivity normalized to $k_{0,th}$, the theoretical conductivity at $R_k=0$, as a function of the characteristic dimensionless parameter $r$ for (a) different volume fractions $V_f$, (b) different aspect ratios $a$, and (c) different fiber orientations governed by the mean angle to heat flux direction $\bar {\theta}$.  The base parameters are $V_f=0.01$, $a=100$, and  $\bar {\theta} =57.3^\circ$. The black dashed line shows the predictions of Eq.~(\ref{eq:four}). }
\end{figure}

Figure~\ref{fig:fig8}%
compares the predictions of Eq.~(\ref{eq:four}) to simulation data collected in this study for different fiber volume fractions, aspect ratios, and orientations. Note that here only networks with fiber mean angles between 57$^\circ$ and 86$^\circ$, i.e., with $\beta>1$, are considered, which is the only regime representative of real-world insulation materials. Figure~\ref{fig:fig8}(a) indicates that the agreement between simulations and theory is improved when the volume fraction increases, as mentioned above. Similarly, Fig.~\ref{fig:fig8}(b) demonstrates a comparable trend concerning the aspect ratio. Note that this parameter has an effect independently of the individual values of $l$ and $d$ and is an important quantity in the conductivity scaling, which is consistent with our previous hypothesis. According to Fig.~\ref{fig:fig8}(c), fiber orientation does not have any effect on the relative variation of $k_{cond,solid}$ with $r$, indicating that the accordance between theory and simulation is fully determined by the values of $V_f$ and $a$. Overall, Fig.~\ref{fig:fig8} shows that the above theoretical model seems to be relevant for 3D fibrous networks in the case of high volume fractions ($V_f>0.05$ when $a= 100$) and high aspect ratios ($a> 450$ when $V_f=0.01$), when adopting representative values of the properties of glasswool for the other parameters.

\vspace{3mm}

\begin{figure}[b]
\includegraphics[width=0.48\textwidth]{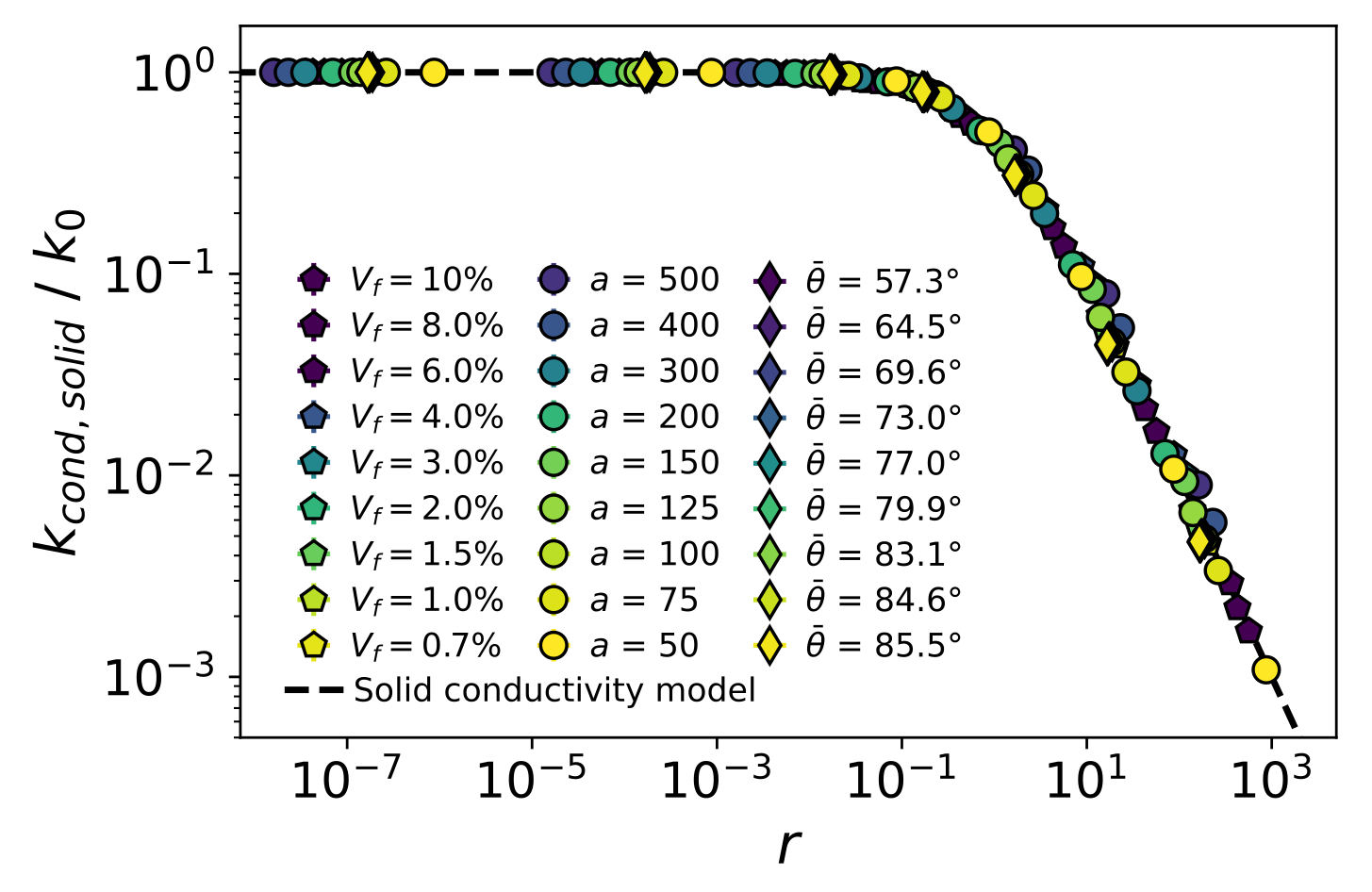}
\caption{\label{fig:fig9} Solid thermal conductivity normalized to $k_0$, the simulated conductivity at $R_k=0$, as a function of the characteristic dimensionless parameter $r$ for different volume fractions $V_f$, aspect ratios $a$, and mean angles $\bar {\theta}$. The base parameters are $V_f=0.01$, $a=100$, and $\bar {\theta} =57.3^\circ$. The black dashed line shows the predictions of Eq.~(\ref{eq:four}). }
\end{figure}

Fig.~\ref{fig:fig9}%
represents, for different fiber parameters, the solid conductivity normalized to $k_0$, its value at $R_k=0$, as a function of the characteristic dimensionless parameter $r$. Simulated data are compared with the prediction $\frac {1}{1+r}$ given by Eq.~(\ref{eq:four}). One can see that all markers corresponding to the simulation data collapse into a single master curve, regardless of the fiber volume fraction, aspect ratio, or mean orientation, which is in good agreement with the theoretical model. This comparison validates, for any value of these network parameters, the consideration of $r$ as the characteristic parameter governing the evolution of the solid conductivity as the contact resistance (or the intrinsic conductivity of fibers) changes. In the general case of a 3D fibrous assembly represented by straight lines, this quantity correctly describes the joint effect of resistance ratio and geometric properties. Obtaining this master curve also indicates that the deviation observed in Figs.~\ref{fig:fig8}(a) and (b) between the model and simulation corresponds to a variable offset in the solid conductivity. Hence, a correction term appears to be needed in the theoretical prediction $k_{0,th}=\frac{1}{4} k_{fib} \pi d^2 \langle n_z \rangle \langle\lvert\cos \theta\rvert\rangle$ to accurately describe fibrous media with low $a$ or $V_f$. Contrary to fiber orientation, these two parameters are found to have a significant effect on the average number of contacts per fiber $\langle N_c \rangle$, indicating that the accordance with the model from Zhao et al.~\cite{zhao_thermal_2018} and Volkov and Zhigilei~\cite{volkov_heat_2012} may be determined by the degree of connectivity of the simulated network. Indeed, when the aspect ratio or the density is low, the connectivity is low and the correlations between fiber temperatures may become non-negligible~\cite{volkov_thermal_2020}. To validate this hypothesis, a modified version of the theoretical framework is derived in Appendix~\ref{sec:appB}, taking into account the contributions of correlations. This new model results in the following expression for the solid conductivity:  
\begin{eqnarray}
k_{cond,solid}=k_{0,th}h \frac{1}{1+r}
\label{eq:five}
\end{eqnarray}
with
\begin{eqnarray}
h=1+ \frac{\langle \delta T_{ij} \rangle}{\nabla T_z \langle H \rangle}.
\label{eq:six}
\end{eqnarray}
Here $\langle \delta T_{ij} \rangle$ represents the average contribution of correlations in the temperature difference at junctions, and $\nabla T_z = \Delta T/L$ is the global temperature gradient imposed in the sample.

\vspace{3mm}

\begin{figure}[b]
\includegraphics[width=0.48\textwidth]{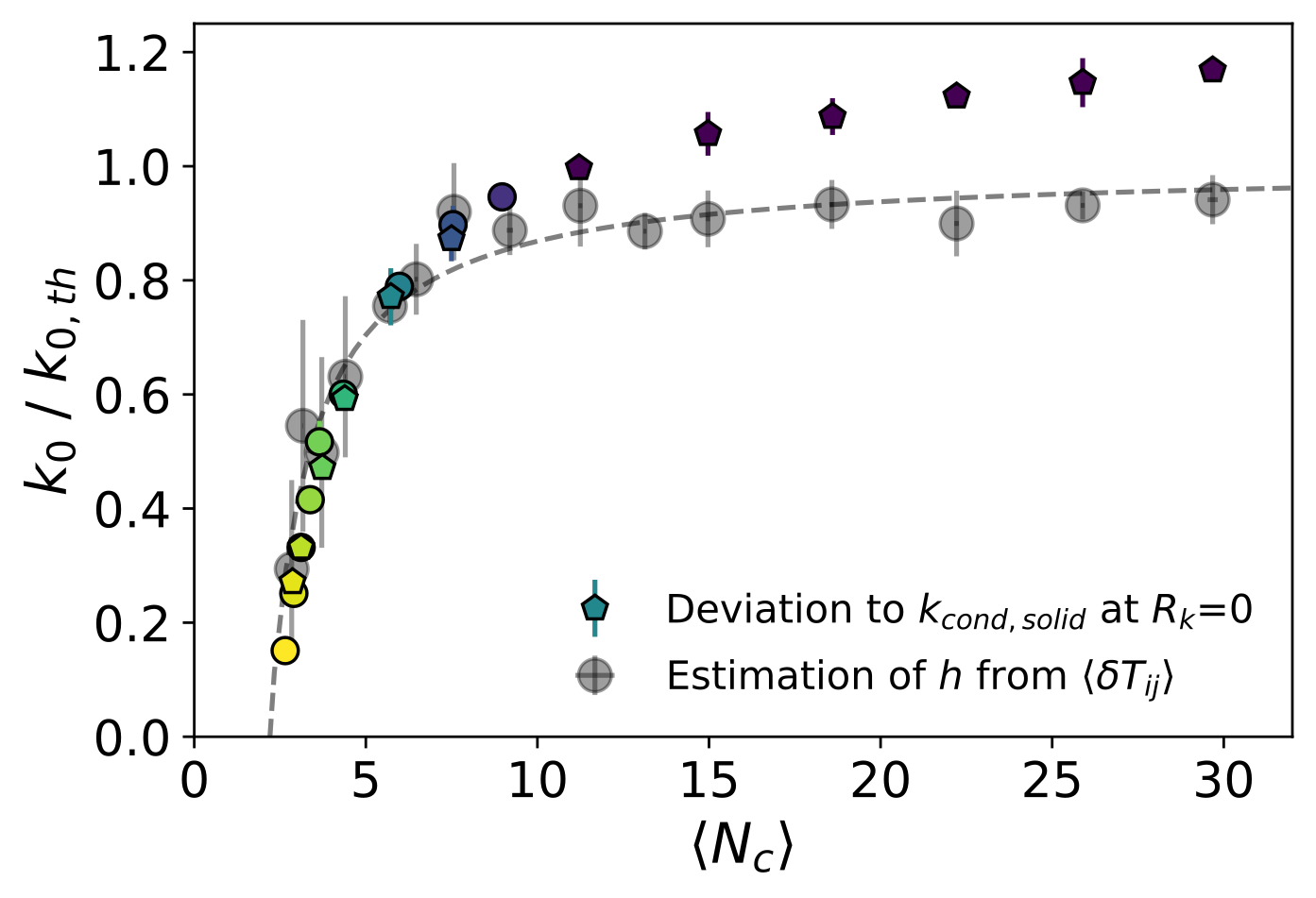}
\caption{\label{fig:fig10} Ratio of $k_0$, the simulated conductivity at $R_k=0$, to $k_{0,th}$, its corresponding theoretical prediction by the initial model according to Eq.~(\ref{eq:four}), as a function of the average number of contacts per fiber $\langle N_c \rangle$. Colored symbols correspond to simulation results presented in Figs.~\ref{fig:fig8}(a) and 8(b), i.e., with different aspect ratios and volume fractions. The gray markers and the associated dashed curve represent the predictions of $h$ calculated from Eqs.~(\ref{eq:six}) and~(\ref{eq:seven}).}
\end{figure}

According to this modified model, the predicted conductivity at any contact resistance is modified by a factor $h$ compared with the initial model; its accuracy is evaluated in Fig.~\ref{fig:fig10}%
.
In this graph, the colored symbols show $k_0/k_{0,th}$, the ratio of the simulated conductivity at $R_k=0$ to its corresponding estimation by the initial model, as a function of $\langle N_c \rangle$. Note that these data correspond to those of Figs.~\ref{fig:fig8}(a) and (b) obtained for different aspect ratios and volume fractions. They show that the evolution of $k_0/k_{0,th}$ can be described by a single curve that depends only on  $\langle N_c \rangle$, independently of the values of $a$ and $V_f$. This finding confirms that the observed reduction in the simulated conductivity compared with that predicted by the initial model is driven by network connectivity, accurately quantified by the average number of contacts per fiber. It can be qualitatively explained by the fact that in a highly connected network, almost all fibers contribute to heat transfer, whereas only a small proportion of them do so in a poorly connected one. Furthermore, the gray symbols in Fig.~\ref{fig:fig10} represent estimations of $h$ obtained by Eq.~(\ref{eq:six}), for different aspect ratios and densities as well. Calculating $h$ only necessitates the additional measurement of $\langle \delta T_{ij} \rangle$ in simulations. This is achieved using the following expression extracted from the derivation of the modified theoretical model (see Appendix~\ref{sec:appB}):
\begin{eqnarray}
\langle \delta T_{ij} \rangle=\langle |\Delta T_{ij}| \rangle - \langle H \rangle \left(\nabla T_z - \biggl\langle \frac{dT}{dz} \biggr\rangle\right)
\label{eq:seven}
\end{eqnarray}
with $\langle |\Delta T_{ij}| \rangle$ the average temperature difference between two fibers at a junction, and $\langle \frac{dT}{dz} \rangle$ the average temperature gradient in fibers along the $z$~direction. The quantity $h$ should correspond to the corrective prefactor $ k_0/k_{0,th}$ needed to account for low-connectivity networks. Figure~\ref{fig:fig10} shows good agreement between these two quantities, despite notable fluctuations in the calculated values of $h$. Note that finding $0\leq h \leq 1$ is a direct result of $\langle \delta T_{ij} \rangle$ being a negative value, indicating that temperature differences at junctions are lower in average in the modified model. Indeed, considering the correlations results in a reduction of temperature differences between closely positioned fibers, especially those in direct contact. A deviation is observed at high $\langle N_c \rangle$, with $ k_0/k_{0,th}$ reaching values higher than 1. As the connectivity increases, the prefactor of the theoretical model should tend towards 1: this suggests that $ k_0/k_{0,th}$ may be slightly overestimated by our simulations, while the estimate for $h$ appears to have rectified this discrepancy, indicating a higher level of accuracy. The dashed line plotted in Fig.~\ref{fig:fig10} represents the best fit of the numerical data obtained for $h$ by a rational function $h=C_1  \frac{\langle N_c \rangle-C_2}{\langle N_c \rangle-C_3}$ where $C_1$, $C_2$, and $C_3$ are adjustable parameters; the choice of such a function is motivated by the form of Eq.~(\ref{eq:six}). This best fit gives $C_1\approx C_3\approx1$ and $C_2\approx 2.18$, confirming that our data are consistent with $h\rightarrow1$  when $\langle N_c \rangle \rightarrow \infty$. The obtained best-fit function can be rewritten as
\begin{eqnarray}
h=1- \frac{\langle N_c \rangle_l -1}{\langle N_c \rangle-1},
\label{eq:eight}
\end{eqnarray}
where $\langle N_c \rangle_l=C_2=2.18$  is the minimum number of contacts per fiber ensuring a conductive network, i.e., that corresponds to the percolation threshold. Note that as the minimum number of contacts for each fiber is 2 to form a fully connected network where thermal transfers can be solved, it was expected that the minimum value averaged at the network scale would indeed be slightly higher. In previous studies~\cite{komori_numbers_1977, volkov_heat_2012}, the average number of contacts per fiber has been expressed as a function of the input parameters: for 3D networks, $\langle N_c \rangle=2aV_f$. Therefore, Fig.~\ref{fig:fig10} can be interpreted as demonstrating percolative behavior for 3D networks governed by the both volume fraction and aspect ratio; in particular, when $a$ approaches 0, regardless of the value of $V_f$, the network conductivity becomes zero, as it falls below the percolation threshold. As stated in Ref.~\cite{volkov_thermal_2024}, this constitutes a major difference with 2D networks, for which the conductivity can remain finite for very small aspect ratios: this is due to the topological difference between 2D and 3D networks.

We can interpret $\langle N_c \rangle -1 $ as the average number of conducting segments per fiber, each one of them being delimited by contact points. Therefore, the deviation from unity of $h$, which is due to the correlation term $\frac{\langle \delta T_{ij} \rangle}{\nabla T_z \langle H \rangle}$ according to Eq.~(\ref{eq:six}), can be entirely explained by this number of segments per fiber. As it decreases and approaches its percolation limit value, temperature distributions along fibers exhibit more pronounced piecewise-linear behavior, therefore increasingly deviating from a linear gradient, as can be seen in Fig. S4 with the Supplemental Material~\cite{supp}. This explains the growing importance of the impact of correlations in the approximate decomposition of the temperature of a fiber made in the theoretical model (see Eq.~(\ref{appk}) in Appendix~\ref{sec:appB}). More generally, this result shows that the solid conductivity in any 3D network of straight fibers can be predicted based on geometric considerations only, using Eqs.~(\ref{eq:five}) and~(\ref{eq:eight}), which do not require resolution of the heat transfer problem.

\vspace{3mm}

Equation.~(\ref{eq:eight}) also allows us to compute a threshold value for $\langle N_c \rangle$, above which the initial model from Zhao et al.~\cite{zhao_thermal_2018} and Volkov and Zhigilei~\cite{volkov_heat_2012} can be considered as sufficient. For a 5\% tolerance, this threshold is $\langle N_c \rangle_{t,5\%}=24.6$. This value may be seen as a separation between two characteristic connectivity regimes for a 3D fibrous assembly, generalizing the low-density and high-density regimes described in Sec. \ref{sec:secIIIA1}, where we gave a rough estimate for a given case with fixed fiber dimensions. In addition, the expression introduced previously linking the average number of contacts per fiber to the network input parameters in 3D networks, $\langle N_c \rangle=2aV_f$, can be used before any numerical generation to estimate the value of $h$ for a given set of network parameters. For an average aspect ratio $a=100$, representative of fibers in glasswool, the threshold $\langle N_c \rangle_{t,5\%}=24.6$ converts into a minimum volume fraction $V_{f_{t,5\%}}=0.123$. This result indicates that to accurately represent most glasswool insulation mats that exhibit lower densities, the model taking into account correlations that we derived and validated in this study should be considered, underlining the interest of this work. 

\vspace{3mm}

\begin{figure}[b]
\includegraphics[width=0.48\textwidth]{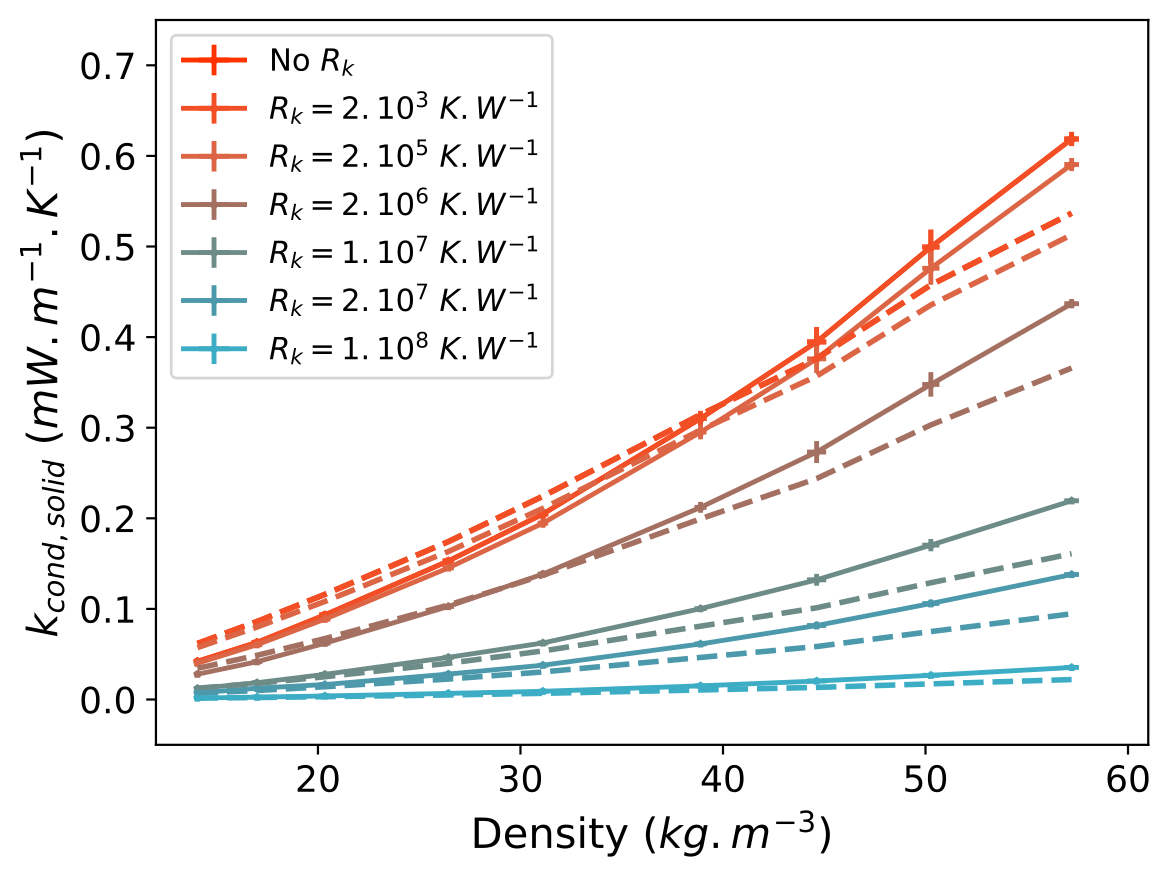}
\caption{\label{fig:fig11} Simulated (plain line) and theoretically estimated (dashed line) solid conductivities as a function of the estimated glasswool density, calculated based on the fiber volume fraction and considering the additional presence of a binder in real samples. The constant input parameters are $l=1$~mm, $d=10~\mu$m, and $\bar {\theta} =79.9^\circ$, with fiber volume fractions $V_f$ ranging from 0.006 to 0.03. The solid conductivities estimated using the theoretical model presented above are obtained using Eqs.~(\ref{eq:five}) and~(\ref{eq:eight}).}
\end{figure}

In practice, to predict the thermal conductivity of a specific fibrous insulation material using this model, according to Eqs.~(\ref{eq:five}) and~(\ref{eq:eight}), the input parameters $d$, $l$, $\langle\lvert \cos\theta\rvert \rangle$, $V_f$, $R_k$, and $k_f$ must be known. The three additional quantities $\langle N_c \rangle$, $\langle H \rangle$, and $\langle n_z \rangle$ necessary to calculate $k_{cond,solid}$ can then be obtained from the numerical generation of a fibrous network characterized by the above parameters, as done in this study. For example, Fig.~\ref{fig:fig11}%
compares our results for $k_{cond,solid}$ calculated by simulations and by this theoretical approach based on morphological parameters only, using as input parameters $d=10~\mu$m, $l=1$~mm, $\bar {\theta}=79.9^{ \circ}$, and $k_f=1.3  $~W~m$^{-1}$~K$^{-1}$, considered, as mentioned before, as relevant average values when considering glasswool. The selected volume fractions, lower than 0.025, typically represent real-world samples, exhibiting densities lower than 60~kg~m$^{-3}$. For each fiber volume fraction, the corresponding material density is estimated from the glass density $\rho_{glass} = 2200$~kg~m$^{-3}$ and taking into account the presence of an organic binder in which the fibers are embedded, with a density twice as low as that of glass and which represents around 10$\%$ of the total solid volume~\cite{noauthor_private_2024}. The values of fiber-to-fiber contact resistances $R_k$, which are not precisely known in real samples, are picked within a probable range, set from 0 to $10^8$~K~W$^{-1}$. Previous studies have indeed estimated the contact resistance between two glass fibers with a diameter of 10~µm to be of the order of $10^7$~K~W$^{-1}$~\cite{perros_caracterisation_2017,nguyen_thermal_2020,doumouro_quantitative_2021}, and it is likely that the binder can only reduce this resistance, by allowing larger contact areas. Figure~\ref{fig:fig11} indicates that the solid conductivity for glasswool densities ranging from 10 to 60~kg~m$^{-3}$ would be lower than 0.7 mW~m$^{-1}$~K$^{-1}$. In the density range studied here, a contact resistance $R_k=10^7$~K~W$^{-1}$ would approximately lead to a $55\%$ to $70\%$ decrease in  $k_{cond,solid}$ compared with the case of infinitely conductive contacts; this impact of contacts tends to increase when the density decreases, as already demonstrated in Sec.~\ref{sec:secIIIA1}. According to Fig.~\ref{fig:fig11}, predicting $k_{cond,solid}$ using the theoretical approach based on the numerical generation of fibrous networks leads to a 13$\%$ maximum relative deviation with simulation results.
An alternative for the prediction of solid conductivity with the theoretical model could also be to measure the needed geometric parameters in reconstructed structures from X-ray tomography images, such as those studied in Ref.~\cite{meftah_multiscale_2019} in the case of glasswool, and to substitute them into Eqs.~(\ref{eq:five}) and~(\ref{eq:eight}).
Note that this theoretical model has been derived for constant values of input parameters, and hence the need to consider average values when applying it to the prediction of solid conductivity of real-world insulation materials.

Experimentally, the thermal performance of insulation materials in the steady-state is usually measured using the guarded hot plate method (ISO 8302) or the heat-flux meter method (ISO 8301): the heat flux going through the material is measured when it is placed between two plates at different temperatures, and the effective conductivity is obtained by Fourier’s law. As stated in the Introduction, radiation and conduction through gas also affect this quantity and cannot be overlooked in this kind of experiment. To validate our modeling results, a crude approach could be to compare the simulated solid conductivity with an estimate derived from measurements of the effective conductivity, after subtracting predicted contributions of radiation and gaseous conduction. Langlais and Klarsfeld~\cite{langlais_isolation_2004} previously estimated the effective conductivity for low densities typical of glasswool mats, between 10 and 60~kg~m$^{-3}$, to be of the order of 30-45 mW~m$^{-1}$~K$^{-1}$. The effective conductivity is measured with an uncertainty of $\pm$ 3$\%$, as requested by ISO standards: this results in a confidence interval covering approximately 1.8 to 2.7 mW~m$^{-1}$~K$^{-1}$, which could even be broadened considering the challenging aspects of accurately predicting the radiative contribution, which depends on the structure of the material and on the levels of temperature involved~\cite{langlais_influence_1995}. Fig.~\ref{fig:fig11} indicates that, when considering input parameters representative of those of glasswool, the simulated solid conductivity is below 0.7 mW~m$^{-1}$~K$^{-1}$. Given this relatively small contribution compared with the uncertainty levels associated with the proposed experimental evaluation procedure, one can only state that the order of magnitude of $k_{cond,solid}$ predicted by our model is consistent with the measured value of the effective conductivity of glasswool mats. Only cryogenic experiments in vacuum aimed at suppressing the contributions of conduction and radiation would allow a proper validation of solid conductivity predictions.

\section{\label{sec:concl}Conclusion}

In summary, we have numerically generated 3D networks representative of fibrous insulation materials such as glasswool, and we have developed an original and efficient nodal simulation approach to analyze their steady-state solid thermal conductivity, integrating the effect of fiber-to-fiber thermal contact resistances. 
We have investigated the effects of fiber length, diameter, volume fraction and angle, and we have demonstrated that the evolution of solid conductivity in 3D networks is governed by a master curve that depends on a characteristic ratio of thermal resistances, which is an innovative way of describing the behavior of this quantity. 
This approach also constitutes an exhaustive validation in the 3D case of the underlying theoretical model, which had previously been numerically validated in the specific case of 2D carbon nanotube networks~\cite{zhao_thermal_2018,volkov_heat_2012,volkov_thermal_2020}; a first comparison with 3D simulation results was presented only recently~\cite{volkov_thermal_2024}, but for more limited ranges of fiber geometric parameters.
In particular, a distinction of this present work is the consideration of various fiber orientations, i.e., the investigation of anisotropic systems, which is particularly important when considering insulation materials like glasswool. 
In this material, fibers also exhibit relatively low aspect ratios and low densities: we have shown that in these cases a modified theoretical model is required to take into account the influence of correlations between fiber temperatures. We have in particular introduced a simple form of this model, applicable to a wide range of fibrous materials, highlighting the significance of the average number of contacts per fiber as a key geometric parameter to describe the degree of connectivity of the networks. 
This model, which enables predictions of solid conductivity based solely on morphological parameters, represents a significant advancement in enhancing the accuracy of thermal performance predictions for insulation materials.

All the conclusions drawn in this study also apply to the prediction of the electrical conductivity of 3D fibrous materials due to the thermal-electric analogy~\cite{verma_thermal_1991}. Therefore, this work paves the way for further investigation of the conduction of heat or electricity through the solid structure in fibrous materials with more complex properties. We plan to extend the use of the nodal method to assemblies of nonstraight fibers, which could be generated using an approach similar to that of Kallel et al.~\cite{kallel_design_2022}. A lower solid conductivity is expected when introducing fiber curviness, as the effective length between the ends of fibers will be reduced, thus decreasing the number of conduction paths~\cite{fata_effect_2020,hicks_effect_2018}. For a more accurate representation of real fibrous samples, the variability of fiber parameters in real samples should also be accounted for, using distributions in the model, as well as the correlations existing between them. The presumed link between contact resistance and fiber diameter was indeed already discussed as an interesting next inclusion in the model, but the link between fiber length and diameter could also be explored. Finally, we plan to integrate in the model presented here the results of future experimental measurements of contact resistance between two glass fibers, in order to provide insight into the extent of the impact of contacts in real glasswool samples.

\begin{acknowledgments}
The authors are grateful to Jean-Marie Chassot (Institut Langevin), who recorded the optical coherence tomography image of a glasswool mat shown in the supplementary material, using LLTech commercial system.  This work pertains to the French government program "France2030"  (EUR INTREE and LABEX INTERACTIFS). This work is supported by the ANRT (Association nationale de la recherche et de la technologie) thanks to a CIFRE fellowship granted to C.G. This work is also supported by the "Investissements d’Avenir" program launched by the French Government (Labex WiFi).
\end{acknowledgments}

\appendix

\section{\label{sec:appA}Resolution of the linear thermal circuit system using a nodal method }

In the steady-state configuration considered in this study, the thermal network can be described by two matrices: the incidence matrix and the conductance matrix. They respectively account for the two factors affecting the overall conductivity: the network structure and the material properties~\cite{kim_systematic_2018}. 

\vspace{4mm}

The incidence matrix of a graph with $n$ nodes and $m$ branches, $\boldsymbol{A}=(a_{fg})_{1\leq f\leq m,1\leq g\leq n}$ is defined by~\cite{dorfler_electrical_2018}
\begin{equation}
a_{fg}= \left\{
    \begin{array}{ll}
        0 & \text{ \footnotesize  if branch $\it{f}$ is not connected to node $\it{g}$,}\\
        +1 & \text{\footnotesize  if the flow through branch $\it{f}$ enters node $\it{g}$,}\\
        -1 & \text{\footnotesize  if the flow through branch $\it{f}$ leaves node $\it{g}$.}
    \end{array}
\right.
\label{appa}
\end{equation}

\vspace{2mm}

\noindent The rows of matrix $\boldsymbol{A}$ correspond to the branches, and the columns correspond to the nodes. As the heat fluxes through branches have defined directions, the network is a directed graph, and hence an oriented incidence matrix.

\vspace{4mm}

The conductance matrix $\boldsymbol{G}=(g_{fh})_{1\leq f,h\leq m}$ is diagonal and is of dimension equal to the number of branches $m$:
\begin{equation}
g_{fh}= \left\{
    \begin{array}{ll}
        R_f^{-1} & \text{ if $f= h$}\\
        0 & \text{if $f \ne h$} \\
    \end{array}
\right.
\label{appb}
\end{equation}
with $R_f$ is the thermal resistance associated with branch $f$. If the branch represents a fiber, its resistance is an internal resistance: $R_{f}= \frac {4l_f}{k_{fib} \pi d^2 }$ with $l_f$ the length of the branch (i.e., the distance between the two associated nodes). If it represents a contact, its resistance is a contact resistance: $R_f=R_k$ with $R_k$ a value we impose for all contacts in the assembly. The introduction of the contact resistance as an effective value is chosen due to the soft-core approach used in this numerical model: there is no dependence on the individual properties of the contacts.

The thermal network has to be solved to access two vectors: the vector of temperatures $\boldsymbol{t}$, of size equal to the number of nodes $n$, and that of heat fluxes $\boldsymbol{q}$, of size equal to the number of branches $m$.

\vspace{4mm}

For any branch $f$, the temperature drop is $e_f= t_{g-1}-t_g  + b_f$, where $t_g$ and $t_{g-1}$ are the temperatures of the connected nodes and $b_f$ is the possible temperature source term. The heat flux through the branch is linked to this temperature difference and to the thermal resistance associated with the branch by Fourier’s Law: $ q_f=e_f / R_f$. Expressed for all branches in matrix form, these two equations become
\begin{equation}
\boldsymbol{e}=-\boldsymbol{At}+\boldsymbol{b}
\label{appc}
\end{equation}
and 
\begin{equation}
\boldsymbol{q}=\boldsymbol{Ge}
\label{appd}
\end{equation}
where $\boldsymbol{e}$ is the $m$-dimensional vector of temperature drops over the thermal resistances.
Combining Eqs.~(\ref{appc}) and~(\ref{appd}) gives
\begin{equation}
\boldsymbol{G}^{-1} \boldsymbol{q} + \boldsymbol{At} = \boldsymbol{b}
\label{appe}
\end{equation}
where $\boldsymbol{b}$ is the $m$-dimensional of temperature sources.

\vspace{4mm}

For any node $g$, the balance of heat rates can be expressed as $\sum_{g} q_g +s_g=0$, where $\sum_{g} q_g$  is the algebraic sum of heat fluxes entering the node and $s_g$ is a possible heat flow source. Expressed for all nodes in matrix form, this equation becomes
\begin{equation}
-\boldsymbol{A}^T \boldsymbol{q}=\boldsymbol{s}
\label{appf}
\end{equation}
where $\boldsymbol{A}^T$ is the transpose of the incidence matrix, and $\boldsymbol{s}$ is the $n$-dimensional vector of heat flow sources.

\vspace{4mm}

Combining the two constitutive equations for conductive heat transfer [Eqs. (A5) and (A6)], we obtain
\begin{equation}
 \boldsymbol{0} = -\boldsymbol{A}^T \boldsymbol{GAt} + \boldsymbol{A}^T \boldsymbol{Gb} + \boldsymbol{s}.
 \label{appg}
\end{equation}
In the network configuration described above, there is no heat flux source (i.e., $\boldsymbol{s}=
\boldsymbol{0}$), only temperature sources at the top and bottom nodes: they are associated with the corresponding branches, to include them in vector $\boldsymbol{b}$. Equation~(\ref{appg}) is solved in order to get the vector of temperatures and that of heat fluxes.

\section{\label{sec:appB}Derivation of the low-connectivity model for solid conduction in fibrous networks}

This model is built on a similar basis to that introduced by Volkov and Zhigilei~\cite{volkov_heat_2012} and Zhao et al.~\cite{zhao_thermal_2018} for 2D and 3D dense networks, and that presented later by Volkov and Zhigilei~\cite{volkov_thermal_2020} for 2D semidilute networks. In this proposed version, a corrective prefactor is introduced to accurately represent 3D fibrous networks with low connectivity, i.e., low density or low aspect ratio. In this case, the temperature gradients in the fibers can vary significantly, largely influenced by their proximity to other fibers.

\vspace{4mm}

In the guarded hot plate configuration, Fourier’s law gives the steady-state heat flux in the vertical direction $Q$ as a function of the thermal conductivity of the material, the established temperature gradient in the vertical direction $\nabla T_z$, and the horizontal cross-sectional area $A_z=L^2$. As the thermal conductivity only accounts for conduction through the solid phase, Fourier’s law directly gives the solid conductivity: $Q=-k_{cond,solid} A_z \nabla T_z$. This also implies that $Q$ is the algebraic sum of all heat flows through the fibers that cross the section $A_z$, which means that it can be expressed by an averaging approach, i.e., $Q=-A_z \langle n_z \rangle \langle Q_i \rangle$, where $\langle Q_i \rangle$ is the average flux crossing one fiber (with $Q_i\geq 0$ if heat flows in the fiber from the end closer to the hot plate to the other end; this fiber direction will be referred to as $w_i$), and $\langle n_z \rangle$ is the average areal number density of fibers through a horizontal section. Note that $\langle Q_i \rangle \geq 0$  on average, which makes $Q$ a negative quantity, consistent with the coordinate system chosen in Fig.~\ref{fig:fig1}(a) and as represented in Fig.~\ref{fig:fig3}. The solid conductivity can therefore be given by
\begin{equation}
k_{cond,solid}=\frac{\langle n_z \rangle \langle Q_i \rangle} {\nabla T_z}.
\label{apph}
\vspace{3mm}
\end{equation}

At the fiber scale, flux $Q_i$ can be expressed by a local Fourier’s law: $Q_i=-k_{fib} A_{fib}  \frac{dT_i}{dw_i}$ with $\frac{dT_i}{dw_i}$ the temperature gradient in direction $w_i$ of fiber $i$, $k_{fib}$ the intrinsic fiber conductivity, and $A_{fib}=\pi d^2/4$ the fiber cross section. With the conductivity and cross section held constant for all fibers in the network, an average form can be written as
\begin{equation}
\langle Q_i \rangle =-k_{fib} A_{fib}  \biggl\langle\frac{dT_i}{dw_i}\biggr\rangle.
\label{appi}
\end{equation}
This approach relies on the assumption of a constant average temperature gradient, which takes into account the contribution of fibers conveying heat flux in the opposite direction to the global steady-state heat flux established in the sample. Introducing the average cosine of the polar angle $\theta$, this temperature gradient can be expressed along the $z$-direction as
\begin{equation}
\biggl\langle \frac{dT}{dz} \biggr\rangle = \frac {-\langle \frac{dT_i}{dw_i}\rangle}{\langle \lvert \cos\theta \rvert \rangle}.
\label{appj}
\end{equation}
Hence, the temperature of fiber $i$ at a given $z$~coordinate is given by the decomposition
\begin{equation}
T_i(z)=T_b+\nabla T_z  z_{c_i}+\biggl\langle \frac {dT}{dz} \biggr\rangle  \left( z-z_{c_i}\right)+\delta T_i (z)
\label{appk}
\end{equation}
where $T_b$ is the source temperature at $z=0$, $z_{c_i }$ is the $z$~coordinate of the center of the fiber, and $\delta T_i$ is a corrective term accounting for the deviation of the real temperature distribution along fiber $i$ from the linear one. In particular, this term can represent the contribution of correlations existing between the temperatures of two fibers in contact. For two fibers $i$ and $j$ in contact, at $z=z_{ij}$ (the position of the contact on fiber $i$) and $z=z_{ji}$ (the position of the contact on fiber $j$), respectively, Eq.~(\ref{appk}) gives

$T_i(z_{ij})=T_b+\nabla T_z  z_{c_i}+\langle \frac {dT}{dz} \rangle  \left(z_{ij}-z_{c_i}\right)+\delta T_i (z_{ij})$ 

\noindent and

$T_j(z_{ji})=T_b+\nabla T_z  z_{c_j}+\langle \frac {dT}{dz} \rangle  \left(z_{ji}-z_{c_j}\right)+\delta T_j (z_{ji})
\vspace{0.7mm}$.

\noindent As the values chosen in this study for fiber diameter, which acts as the distance threshold for contact detection, are about 2 orders of magnitude smaller than the fiber length, it is reasonable to assume that $z_{ij} \approx z_{ji}$. Then, the temperature difference at a junction is given by
\begin{multline}
\Delta T_{ij}= T_j(z_{ji})-T_i(z_{ij})\\
=
\left(\nabla T_z - \biggl\langle \frac {dT}{dz} \biggr\rangle\right)\left(z_{c_j}-z_{c_i}\right)+ \delta T_{ij}
\label{appl}
\vspace{0.5mm}
\end{multline}
\noindent where $\delta T_{ij}=\delta T_j (z_{ji})-\delta T_i (z_{ij})$ represents the correction on the estimated temperature drop due to correlations. Considering heat transfer will occur from the fiber whose center is closer to the hot plate to the other one, and introducing $\langle H \rangle = \lvert z_{c_j}-z_{c_i} \rvert $ as the average $z$~distance between two fibers in contact, averaging Eq.~(\ref{appl}) results in
\begin{equation}
\langle \lvert \Delta T_{ij} \rvert \rangle = \langle H \rangle \left(\nabla T_{z}-\biggl\langle \frac {dT}{dz} \biggr\rangle\right) + \langle \delta T_{ij} \rangle.
\label{appm}
\vspace{2mm}
\end{equation}

On average, one can assume that the flux through one fiber $\langle Q_i \rangle$ flows in or out through half of the contacts, as each contact corresponds to two fibers: $\langle Q_i \rangle = \langle \frac {1}{2}  \sum_j \lvert Q_{ij} \rvert \rangle $, with $\lvert Q_{ij} \rvert$ the net interface flux between fibers $i$ and $j$, governed by the value of the contact resistance, i.e., $\lvert Q_{ij} \rvert = \frac  {\lvert T_j-T_i \rvert}{R_k}$.
This flow balance can be therefore written as
\begin{equation}
\langle Q_i \rangle= \frac {\langle N_c \rangle \langle \lvert \Delta T_{ij} \rvert \rangle} {2R_k}
\label{appn}
\vspace{1mm}
\end{equation}

\noindent where $\langle N_c \rangle$ is the average number of contacts per fiber. Using Eqs.~(\ref{appi}),~(\ref{appj}) and~(\ref{appn}), the following relationship between the average temperature gradient along fibers in the $z$~direction and the average temperature drop at a junction is obtained:
\begin{equation}
\biggl\langle \frac {dT}{dz} \biggr\rangle = \frac {\langle Q_i \rangle} {\langle \lvert \cos\theta \rvert \rangle k_{fib} A_{fib}} = \frac {\langle N_c \rangle \langle \lvert \Delta T_{ij} \rvert \rangle} {2 R_k \langle \lvert \cos\theta \rvert \rangle k_{fib} A_{fib}}.
\label{appo}
\end{equation}
By substituting Eq.~(\ref{appo}) into Eq.~(\ref{appm}), one can find the following expression for the average temperature drop at junctions:
\begin{equation}
\langle \lvert \Delta T_{ij} \rvert \rangle = \nabla T_z \langle H \rangle \frac {1+ \frac {\langle \delta T_{ij} \rangle}{\nabla T_z \langle H \rangle}}{1+ \frac {\langle H \rangle \langle N_c \rangle}{2 R_k \langle \lvert \cos\theta \rvert \rangle k_{fib} A_{fib}}}.
\label{appp}
\vspace{4mm}
\end{equation}

Combining Eq.~(\ref{apph}) and Eq.~(\ref{appn}), we have
$k_{cond,solid}=
\frac {\langle n_z \rangle  \langle N_c \rangle \langle \lvert \Delta T_{ij} \rvert \rangle}{2 R_k \nabla T_z}$. This expression for the solid conductivity can be inserted into Eq.~(\ref{appp}), which gives its evolution as a function of $\nabla T_z$, $\langle \delta T_{ij} \rangle$, and the geometric parameters of the network:
\begin{multline}
k_{cond,solid}= k_{fib} A_{fib} \langle n_z \rangle \langle \lvert \cos\theta \rvert \rangle \left(1+ \frac {\langle \delta T_{ij} \rangle}{\nabla T_z \langle H \rangle}\right)\\
\times
\frac{1}{1+ \frac{2 R_k \langle \lvert \cos\theta \rvert \rangle k_{fib} A_{fib}}{\langle H \rangle \langle N_c \rangle}}.
\label{appq}
\vspace{2mm}
\end{multline}

\noindent Using the notation introduced in Sec.~\ref{sec:secIIIB},
$r=\frac {2l\langle \lvert \cos\theta\rvert\rangle}{\langle H \rangle \langle N_c \rangle}  \frac {R_k}{R_{fib}} = \frac {2R_k \langle \lvert \cos \theta \rvert \rangle k_{fib} A_{fib}} {\langle H \rangle \langle N_c \rangle}$  and $k_{0,th}=k_{fib} A_{fib} \langle n_z \rangle \langle \lvert \cos \theta \rvert \rangle $, Eq.~(\ref{appq}) can be written as
\begin{equation}
k_{cond,solid}=\left(1+\frac {\langle \delta T_{ij} \rangle} {\nabla T_z \langle H \rangle}\right) k_{0,th} \frac {1}{1+r}.
\label{appr}
\end{equation}

\bibliography{bib_article}

\end{document}


\title{Supplementary Material for the manuscript:\\ Modelling conductive thermal transport in three-dimensional \\fibrous media with fiber-to-fiber contacts}

\author{Clémence Gaunand}
\affiliation{
 Institut Langevin, ESPCI Paris, Université PSL, CNRS - Paris (France),
}%
\affiliation{
 Saint-Gobain Research Paris - Aubervilliers (France)
}%
\affiliation{
 Institut Pprime, CNRS, Université de Poitiers, ISAE-ENSMA - Poitiers (France)
}%

\author{Yannick De Wilde}
\affiliation{
 Institut Langevin, ESPCI Paris, Université PSL, CNRS - Paris (France),
}%

\author{Adrien François}
\affiliation{
 Saint-Gobain Research Paris - Aubervilliers (France)
}%

\author{Veneta Grigorova-Moutiers}
\affiliation{
 Saint-Gobain Research Paris - Aubervilliers (France)
}%

\author{Karl Joulain}
\affiliation{
 Institut Pprime, CNRS, Université de Poitiers, ISAE-ENSMA - Poitiers (France)
}%

\date{\today}
\maketitle

\section*{}
\renewcommand{\thefigure}{S\arabic{figure}}

\begin{figure}
\includegraphics[width=0.7\textwidth]{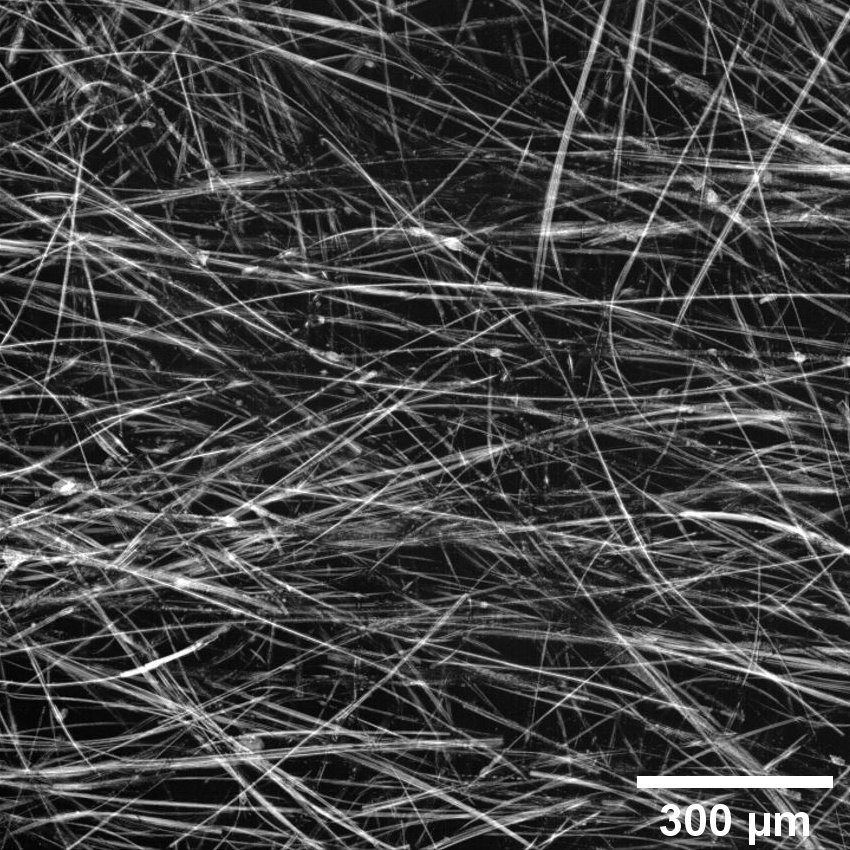}
\caption{\label{fig:figs1} Optical coherence tomography image of a glasswool mat (top view), with micro-scale resolution, showing the low tortuosity of fibers in this material.}
\end{figure}

\begin{figure}
\includegraphics[width=0.6\textwidth]{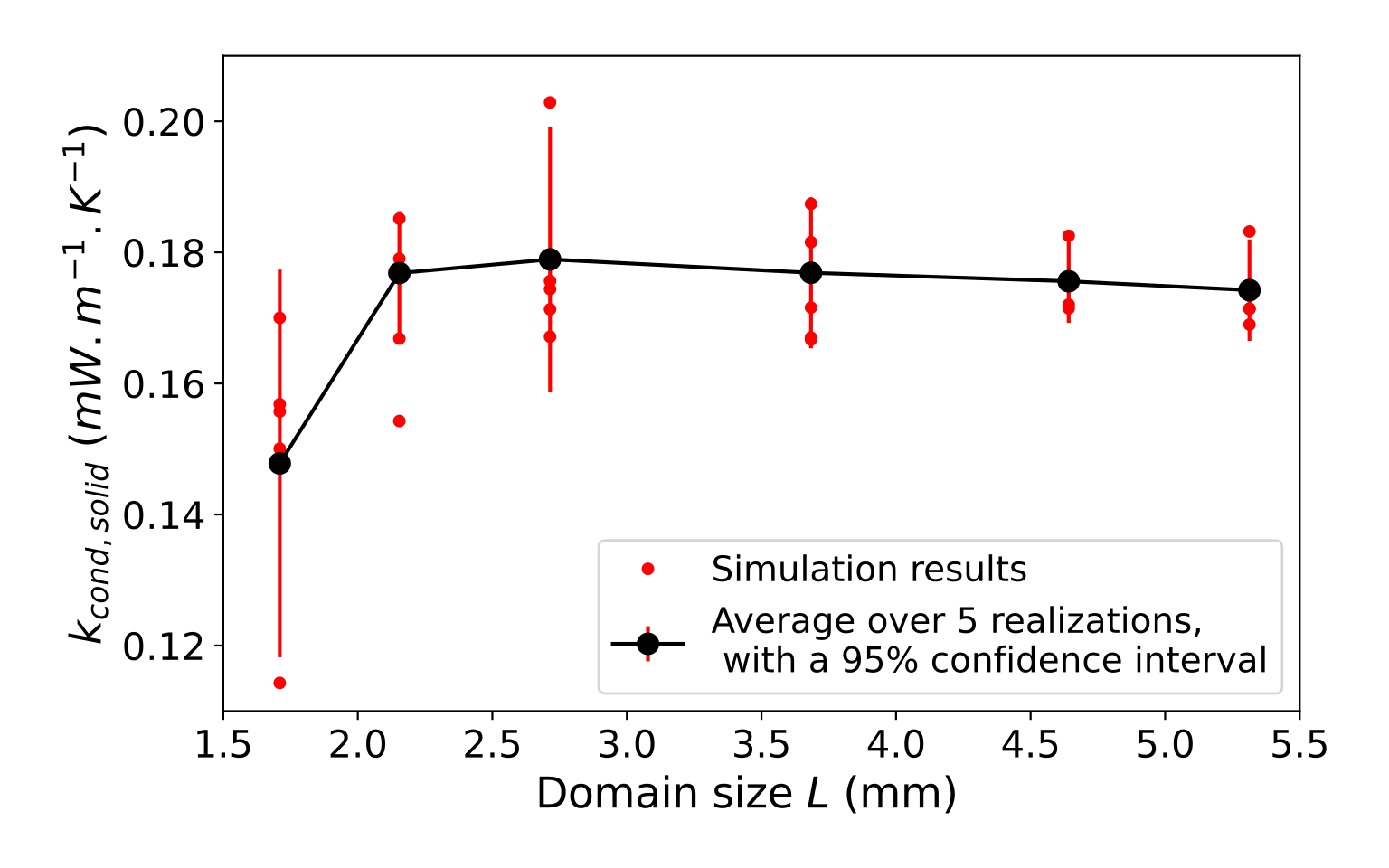}
\caption{\label{fig:figs2}  Illustration of the Representative Volume Element (RVE) selection procedure: for each set of input parameters, the domain size is increased until convergence of the simulation results, which is quantitatively identified by a relative error on the solid thermal conductivity averaged over 5 realizations at a $95\%$ confidence level $\epsilon_{rel}$ lower than 5$\%$. In the case shown here, the input parameters are $V_f=0.01$, $l=1$~mm, $d=10~\mu$m and $\bar{ \theta}=79.9^\circ$, and the RVE determined with this method is $91.125$~mm$^{3}$~($L=4.5$~mm).}
\end{figure}

\begin{figure}
\includegraphics[width=1\textwidth]{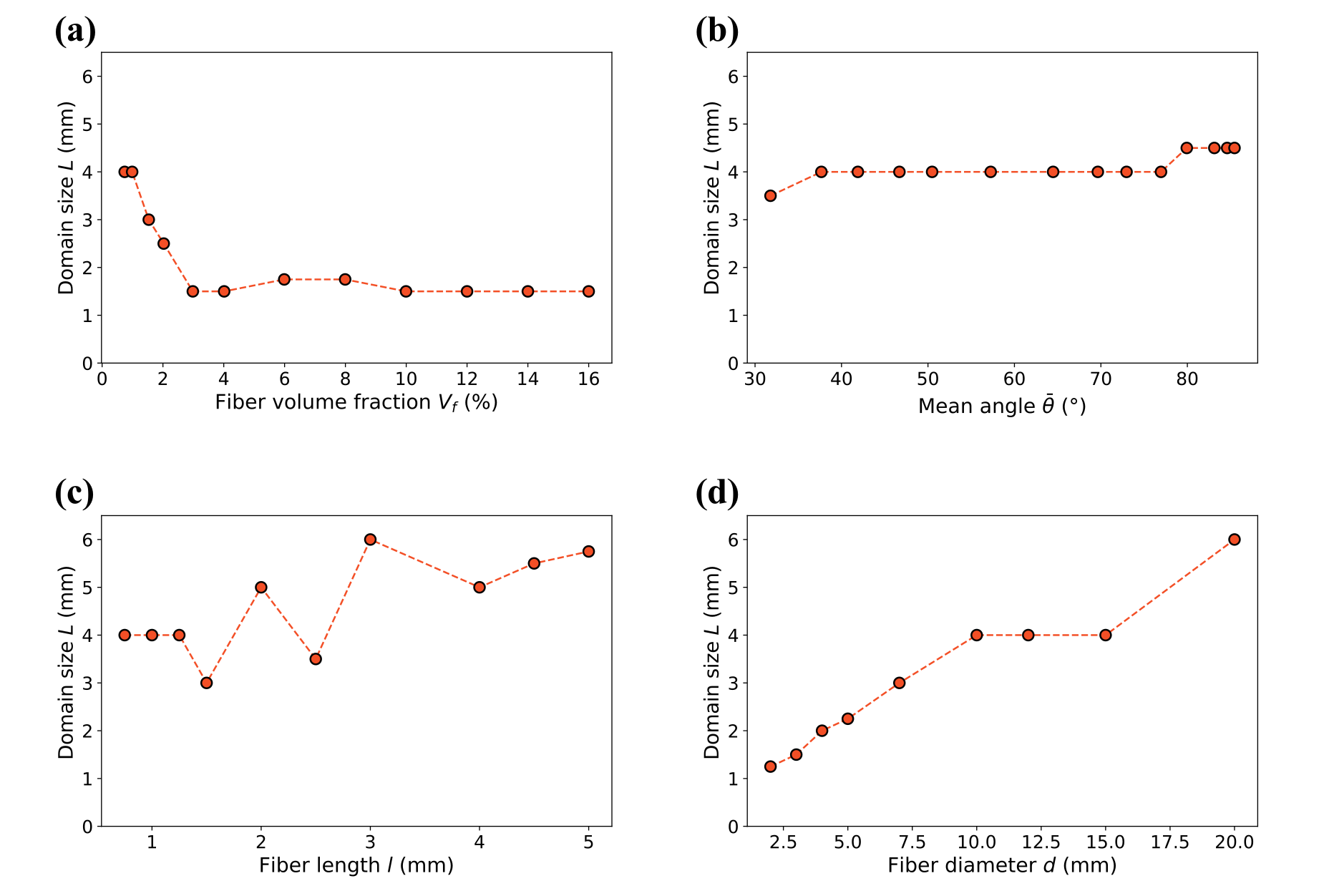}
\caption{\label{fig:fig3}  Domain sizes $L$ used in the simulations presented in this paper, corresponding to a RVE in each case: (a) for different volume fractions, corresponding to data shown in Fig. 4, (b) for different fiber orientations, corresponding to data shown in Fig. 5, (c) for different fiber lengths, corresponding to data shown in Fig. 6, and (d) for different fiber diameters, corresponding to data shown in Fig. 7. The base parameters are $V_f=0.01$, $\bar {\theta} =57.3^\circ$, $l=1$~mm and $d=10~\mu$m.}
\end{figure}

\begin{figure}
\includegraphics[width=0.52\textwidth]{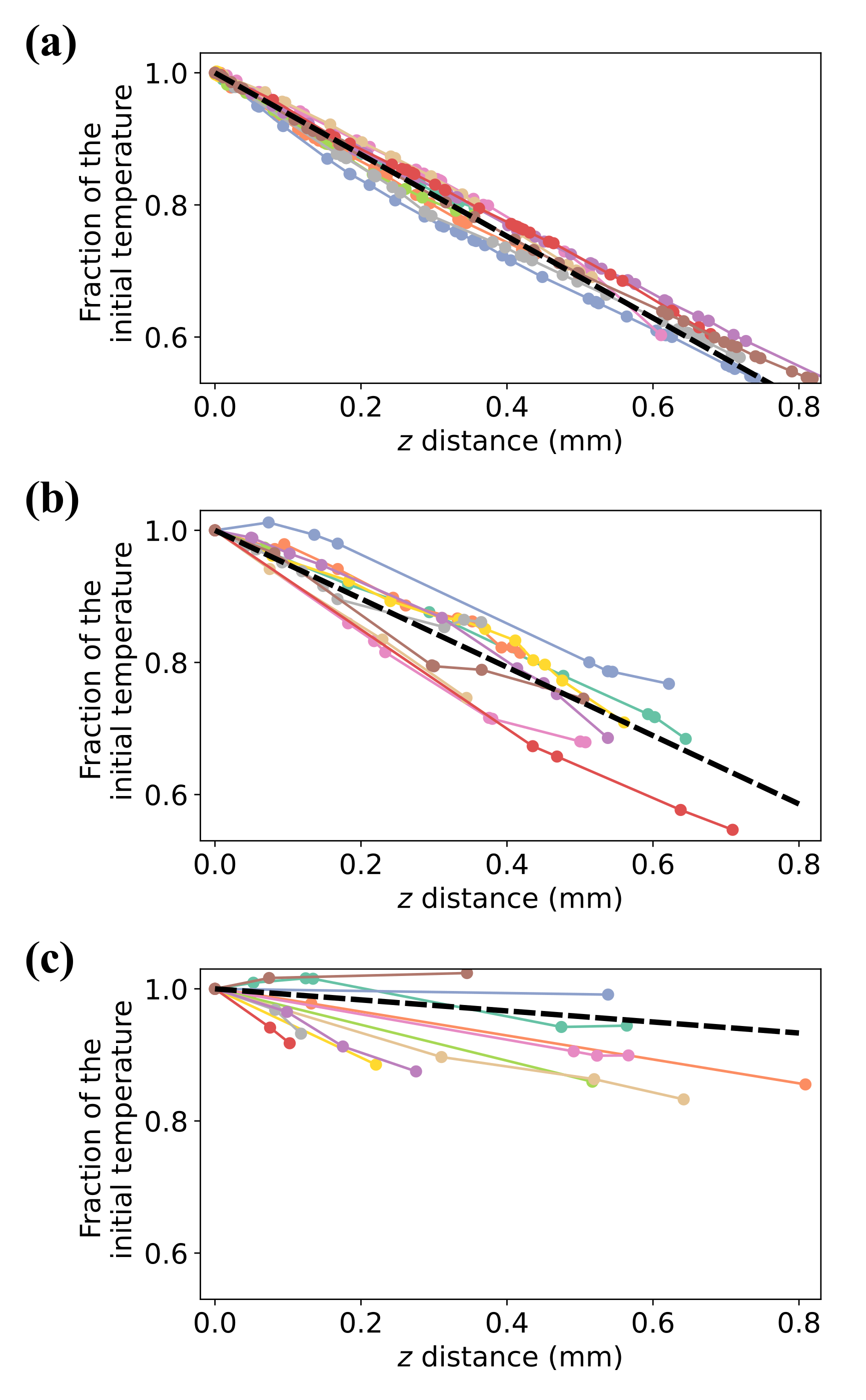}
\caption{\label{fig:figs4} Temperature distributions along different representative fibers as a function of the vertical position (in color), compared in relative variation to the average linear temperature gradient in fibers in the $z$ direction $\langle \frac{dT}{dz} \rangle$ (slope of the dashed black curve), in networks with (a) $\langle N_c \rangle=22.2$, (b) $\langle N_c \rangle$ = 5.7, and (c) $\langle N_c \rangle$~=~3.1.}
\end{figure}